\documentclass[reprint,prb,amsmath,groupedaddress,amssymb,floatfix,nofootinbib,bibnotes,longbibliography,superscriptaddress,twocolumn]{revtex4-2}

\usepackage{amsmath,amsfonts,amssymb}
\usepackage{mathrsfs}
\usepackage{graphicx}
\usepackage[english]{babel} 
\usepackage{verbatim}
\usepackage[colorlinks=true,citecolor=blue,linkcolor=magenta]{hyperref}
\usepackage[usenames, dvipsnames]{color}
\usepackage{bbold}
\usepackage{chemformula}
\usepackage{braket}
\usepackage{dsfont}

\newcommand{\hc}{\mathrm{h.c.}} 

\begin{document}

\title{A solvable model for strongly interacting nonequilibrium excitons}

\author{Zhenhao Song}
\thanks{These authors contributed equally to this work}
\affiliation{Physics Department, University of California, Santa Barbara, California 93106-4030, USA}

\author{Tessa Cookmeyer}
\thanks{These authors contributed equally to this work}
\affiliation{Kavli Institute for Theoretical Physics, University of California, Santa Barbara, California 93106-4030, USA}

\author{Leon Balents}
\affiliation{Kavli Institute for Theoretical Physics, University of California, Santa Barbara, California 93106-4030, USA}
\affiliation{Canadian Institute for Advanced Research, Toronto,  Ontario, Canada}
\affiliation{French American Center for Theoretical Science, CNRS, KITP, Santa Barbara, California 93106-4030, USA}

\begin{abstract}
We study the driven-dissipative Bose-Hubbard model with all-to-all hopping and subject to incoherent pumping and decay, as is naturally probed in several recent experiments on excitons in \ch{WS2}/\ch{WSe2} moir\'e systems, as well as quantum simulators. By positing a particular form of coupling to the environment, we derive the Lindblad jump operators and show that, in certain limits, the system admits a closed-form expression for the steady-state density matrix. Away from the exactly solvable regions, the steady-state can be obtained numerically for 100s-1000s of sites. We study the nonequilibrium phase diagram and phase transitions, which qualitatively matches the equilibrium phase diagram, agreeing with the intuition that increasing the intensity of the light is equivalent to changing the bosonic chemical potential. However, the steady-states are far from thermal states and the nature of the phase transitions is changed.
\end{abstract}

\maketitle

\section{Introduction}

The Bose-Hubbard model is the paradigmatic model of interacting bosons on a lattice and describes a transition between a Mott insulating and superfluid phase \cite{Fisher1989}. There are now several platforms for experimentally realizing this model such as trapped bosonic atoms \cite{takasu2020energy,Endres2011,yang2020observation,Su2023,greiner2002quantum} or photon modes \cite{noh2016quantum,Carusotto2013,ballarini2019polaritonics}. In the latter case, interactions are induced  through circuit quantum electrodynamics (QED) \cite{houck2012chip,Raftery2014}, cavity QED \cite{Hoffman2011,fink2008climbing,Spillane2008,Fitzpatrick2017}, semiconductor platforms \cite{fink2018signatures,Rodriguez2017,Goblot2019}, or Rydberg atoms \cite{Gorshkov2011,peyronel2012quantum}, and a lattice can be produced by interconnecting these platforms \cite{Goblot2019,Fitzpatrick2017,Carusotto2009}.
Excitons in transition-metal dichalcogenide moir\'e bilayers \cite{montblanch2021confinement,rivera2015observation, Camacho-Guardian} or 
artificial lattices
\cite{lagoin2022extended} provide another emerging solid-state realization of Bose-Hubbard physics,
where the interactions are naturally implemented through strong dipole repulsions.
All these experimental settings move beyond the study of equilibrium physics as these systems either controllably couple to their environments or have finite lifetimes for the bosons \cite{muller2012engineered,houck2012chip}.

We are most interested in understanding several recent experiments potentially realizing bosonic Mott physics with excitons in \ch{WSe2}/\ch{WS2} moir\'e bilayers \cite{xiong2023correlated,park2023dipole,lian2024valley,gao2024excitonic,xiong2024tunable}. The experiments see a finite energy shift in photoluminescence spectrum interpreted as an on-site exciton-exciton interaction with multiple excitons per site. 
There is considerable evidence that the excitons enter an interaction-dominated Mott-insulating regime at sufficiently large intensities of light \cite{xiong2023correlated,gao2024excitonic}. 
The inherently non-equilibrium nature of these experiments, given the necessity of continuously pumping the system full of decaying excitons,  naturally raise the question of how much these reported nonequilibrium Mott phases resemble their equilibrium counterpart, 
and whether an analog of the superfluid phase exists out of equilibrium.

The Bose-Hubbard model,
\begin{equation} \label{eq:H0_full}
    H_0=\mu \sum_{i} n_i + \frac{U}2 \sum_i n_{i}(n_i-1) -\sum_{i,j} t_{i,j} b_i^\dagger b_j^{\vphantom{\dagger}},\\
\end{equation}
subject to dissipation and driving, is 
the prototypical model for studying strongly interacting open quantum systems. The drive is often coherently added to the total Hamiltonian, $H_\text{tot} = H_0 + F b_i^\dagger + F^* b_i$,
and the decay is implemented through a Lindblad master equation for the density matrix $\rho$ \cite{Sieberer2014,Biella2017,Lebreuilly2016},
\begin{equation}\label{eq:LME}
    \dot \rho = -i[H_0,\rho] +\sum_{k,k'} \gamma_{k,k'} \mathcal L(L_k,L_{k'}^\dagger)[\rho]
\end{equation}
where $\mathcal L(A,B)[\rho(t)] = A \rho(t) B - \frac12\{BA,\rho(t)\}$, $L_k$ are the jump operators and $\gamma_{k,k'}$ are the associated damping rates. By including suitable Lindblad jump operators, we can include incoherent driving as the reverse process of decay. 

This system has been numerically studied with numerous methods: mean-field theory (MFT)\cite{Caleffi2023,LeBoite2013,LeBoite2014,Wilson2016,Savona2017,Biondi2017,biondi2017spatial}, cluster MFT\cite{Tomadin2010,Huybrechts2020}, the positive $P$ representation\cite{LeBoite2014,Deuar2021}, the corner-space renormalization group method\cite{Finazzi2015}, functional renormalization group using a Keldysh path integral\cite{Sieberer2013,Sieberer2014}, quantum trajectories\cite{Wilson2016,Huybrechts2020,Vicentini2018}, matrix-product-state based methods \cite{MendozaArenas2016,jaschke2018one,Weimer2021} and a truncated hierarchy of correlators \cite{casteels2016truncated}. These studies all assume a simple and decoupled form of Lindblad jump operators, i.e.~$L_{i,-}=b_i$, which works well for describing nearly isolated non-interacting atoms. 
In general, however, such an assumption does not hold for a strongly correlated system when $\gamma_{k,k'}/U \ll 1$.

In this work, we emphasize that, the explicit form of jump operators is determined by the system Hamiltonian itself, as well as the coupling with the environment, which can significantly modify the Lindblad master equation and the properties of the steady state. 
In order to fully understand the phases and dynamics in these nonequilibrium systems, it is useful to have prototypical solvable models \cite{Roberts2023}.
Here we study the Bose Hubbard Hamiltonian \eqref{eq:H0_full} with an all-to-all hopping for $N$ sites, i.e.~ $t_{ij}=t/N$ for any $i,j$. The all-to-all hopping mimics the effect of a mean-field analysis at the equilibrium level, and allows us to move beyond mean-field with analytical control. 
Remarkably, in such an approach we are able to derive the explicit form of the Lindblad jump operators, analytically obtain closed-form solutions of the steady state in certain limits, and numerically solve the system for up to thousands of sites in others.

\begin{figure}[tb]
\includegraphics[width=0.5\textwidth]{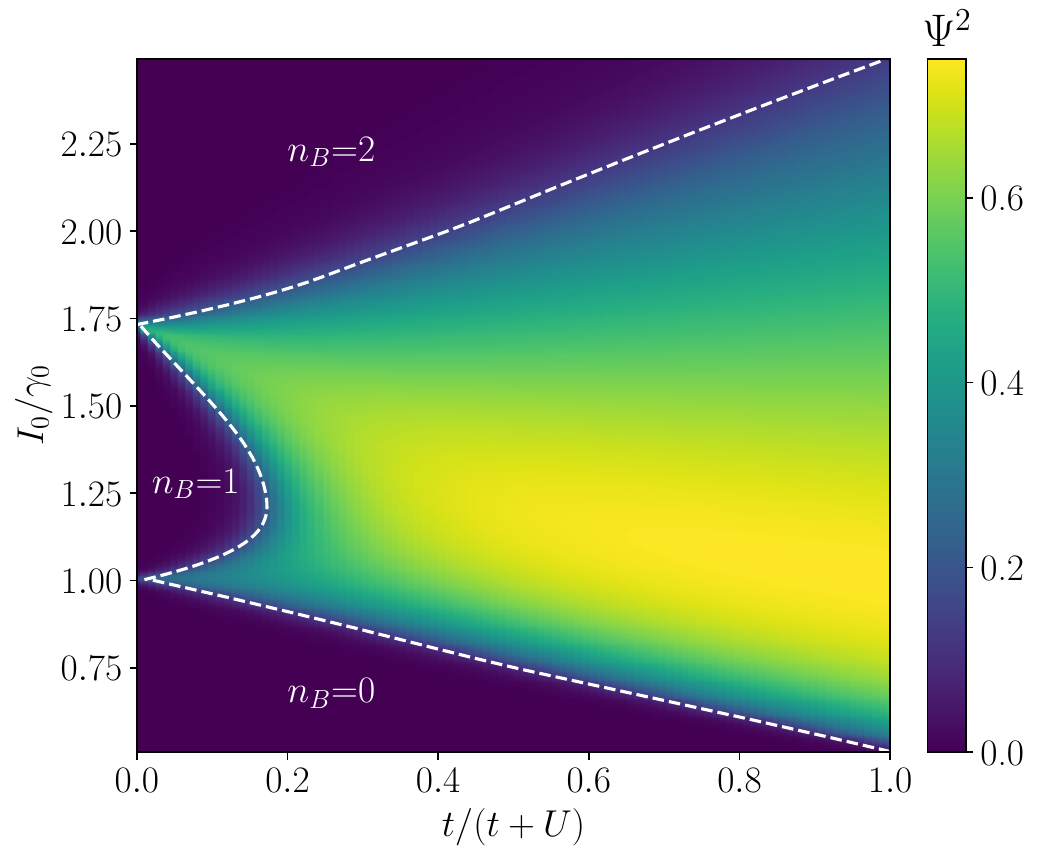}
    \caption{We plot the order parameter $\Psi^2=\sum_{i,j}\langle b_i^\dagger b_j\rangle/N^2$ computed for the steady-state density matrix $\rho_{ss}$, extrapolated to the  $N\to \infty$ limit as a function of the ratio of exciton production and decay rates, $I_0/\gamma_0$, and $t/(t+U)$. 
    $I_0(\gamma_0)$ are pumping(decay) rates at energy $\omega=\mu$. Damping rates at other energies are specified as in Sec.~\ref{sec:numer}.
    The phase boundaries are estimated using the entropy per site $S=-\text{Tr}[\rho_{ss} \ln(\rho_{ss})]/N$, which is maximal at the phase transition. 
    We see that $n_B=0, 1, 2$ in the lobes as indicated on the plot in the SI Appendix.  
}
\label{fig:pd_main}
\end{figure}

Our main results are summarized in the phase diagram, Fig.~\ref{fig:pd_main}, as a function of the ratio of exciton production and decay rates $I_0/\gamma_0$ and $t/(t+U)$.  Similar to the equilibrium model, we see that there are two phases: Mott-insulator phases with an integer values of $n_B=(1/N) \sum_i \langle b_i^\dagger b_i\rangle$ and a superfluid phase identified as a non-zero value of $\Psi^2 = (1/N^2) \sum_{i,j}\langle b_i^\dagger b_j\rangle$ with a generic value of $n_B$.
Although the steady-state density matrices are not of a (thermal) equilibrium form, the nonequilibrium and equilibrium phase diagram resemble each other lending credence to the qualitative understanding of the experiment as tuning the boson chemical potential. We additionally study the critical properties of the phase transition. Notably, the Mott-Mott phase transition, which is simply a level crossing at the equilibrium level, becomes a continuous phase transition and the critical exponents for the Mott-superfluid transition are  distinct from their equilibrium values.

\section{Derivation of the mater equation}

We will analyze the Bose-Hubbard Hamiltonian, \eqref{eq:H0_full} with all-to-all hopping between the $N$ sites, $t_{i,j}=t/N$, and
in the presence of coupling to the  environment. We will assume that $\mu \gg U,t$ as is relevant for the physical system, where $\mu$ is the excitation gap of excitons.

We now proceed to derive the Lindblad master equation in the weak-coupling limit \eqref{eq:LME} for this model. We follow the standard procedure \cite{manzano2020short,breuer2002theory}: 
first, we start with a general coupling form between the system and environment $H_I=\sum_{\alpha}S_{\alpha}\otimes\mathcal{R}_{\alpha}$, 
where $S_{\alpha}$ ($\mathcal{R}_{\alpha}$) are Hermitian operators that act on the system (environment), and the index $\alpha$ labels different operators.
Then, we decompose $S^{\alpha}=\sum_j S_{\alpha;j}$ by eigenoperators of the system Hamiltonian $H_0$, which are defined by the relation $[H_0,S_{\alpha;j}] = -\omega_j S_{\alpha;j}$.
The Lindblad jump operators are nothing but the eigenoperators $S_{\alpha;j}$, and the damping rate $\gamma_{\alpha;j}$ is given by 
$\int_{-\infty}^{+\infty}ds e^{i\omega_j s} \langle \mathcal{R}^{\dagger}_{\alpha}(s) \mathcal{R}_{\alpha}(0)\rangle$.
With these operators in hand, the equation of motion for the system's density matrix can be written out explicitly according to \eqref{eq:LME}.

Since the excitons emit light when they decay, we know that they interact with the quantized electromagnetic field. A typical dipole interaction takes the form $-\sum_i\vec{D}_i\cdot \vec{E}_i$, with $\vec{D}_i\sim (b_i^{\dagger}+b_i^{\vphantom{\dagger}})$. In momentum space, after the rotating-wave approximation, it can be expressed as \cite{Hanai2018}

\begin{equation}
\begin{aligned}
  \sum_{\pmb k,\pm} g_{\pmb k,\pm} a_{\pmb k,\pm} b_{\pmb k}^\dagger
  &\approx \frac{1}{\sqrt{N}}\sum_{\pmb k,\pm} g_{\pmb k,\pm} a_{\pmb k,\pm} \sum_i b_{i}^\dagger
  \label{eq:dipole coupling}
\end{aligned}
\end{equation}
where $g_{\pmb k,\pm}$ is related to the polarization of the light mode. We used $\pmb k \cdot \pmb r_i\ll 1$, since the relevant wavelength of light ($\approx 700$ nm) is much larger than the distance between moir\'e unit cells ($\approx7.5$ nm)\cite{wang2022light}.
Therefore, as opposed to the common assumption where each site couples to different modes in the bath, we assume the environment couples with the system in a collective way, reminiscent of the Dicke model \cite{kirton2019introduction}.
Note that $g_{\pmb k}\sim \mathcal O(\sqrt{N/V})$ where $V$ is the volume available to the photon modes, and we suppress the light's polarization for simplicity.
With such assumption, we may write the coupling as
\begin{subequations}
\begin{align}
    H_I &= \mathcal B^{\dagger} \mathcal R^{\vphantom{\dagger}}  + \mathcal R^\dagger \mathcal B^{\vphantom{\dagger}}
    \\
    \intertext{with}
    \mathcal B&\equiv \sum_i b_i
\end{align}
\end{subequations}
where $\mathcal R$ denotes the environment operator. Though we only argued above for the decay of excitons to have such a form, treating the production and decay on equal footing allows us to make theoretical progress.

 We do not specify the explicit forms of $\mathcal R$ since they could be complicated in general. Specifically, the long-lived interlayer excitons which form the exciton lattice are generated through at least a two-step process, where intralayer excitons are photoexcited and then converted to interlayer excitons through some relaxation process\cite{rivera2015observation}. We therefore consider only the incoherent contributions of $H_I$ to the exciton system. 

We now decompose $\mathcal B$ in $H_I$ in terms of eigenoperators, which, in general, requires diagonalizing $H_0$. 
Note that the coupling to the environment and the Hamiltonian itself are invariant under all permutations of the sites;  we also assume the system's initial state (before turning on exciton-producing effects) is the vacuum $\rho = |0\rangle \langle 0|$ since $\mu>0$, which has permutation symmetry as well.
Therefore, the only states that are accessed are in the fully symmetric sector of this permutation symmetry and it is sufficient to diagonalize $H_0$ in this sector.

We parametrize the sector with the normalized states
\begin{equation}
    |\vec n\rangle = \frac{1}{\sqrt{N! \prod_i (n_i)!}} \sum_{\sigma \in S_N} \sigma |(n_0,n_1,...,n_M)\rangle
\label{eq:nvec}
\end{equation}
where $\vec n$ is a vector in $(M+1)$ dimensional space, $S_N$ is the symmetric group, $\sigma$ is the permutation operator that shuffles site $i$ to $\sigma(i)$, and the state $|(n_0,n_1,...,n_M)\rangle$ has the first $n_0$ sites being unoccupied, the next $n_1$ sites having one boson occupied, etc. We have chosen a maximum number of allowed bosons per site, $M$, and $\sum_i n_i = N$. 

It is convenient to define transition operators $\mathcal O_{n,m} \equiv \sum_i (|n\rangle \langle m|)_i\otimes \prod_{j\neq i} \mathds{1}_j$
which changes site $i$ from having $m$ bosons to $n$ bosons and acts as the identity on all other sites.
Note that $ \mathcal O_{n,m} \sigma = \sigma\mathcal O_{n,m}$ for all $\sigma \in S_N$. We can then find
\begin{equation}\label{eq:O_act_n}
    \mathcal O_{n,m} |\vec n\rangle =|\vec n - \hat e_m + \hat e_n\rangle  \begin{cases}
        \sqrt{ n_m (n_n+1)} & \text{if $n\ne m$} \\
        n_m & \text{if $n=m$}
    \end{cases}
\end{equation}
where $\hat e_i$ are the unit vectors and $n_m$ are the components of $\vec n$.

Letting $\hat N_B$ be the total boson number operator with eigenvalues $N_B$, we use these relations to write $H_0$ in the $|\vec n\rangle$ basis:
\begin{equation}
\begin{aligned}\label{eq:H0eigprob}
    &H_0|\vec n\rangle = (\mu N_B + \frac{U}{2} \sum_n n (n-1) n_n)|\vec n\rangle \\
    &- \frac{t}{N}\sum_{nm}\sqrt{nm n_m n^{(m)}_{m-1} n^{(m)}_{n-1}(n^{(m)}_n+1)} |\vec n^{(m)} - \hat e_{n-1} + \hat e_n\rangle 
\end{aligned}
\end{equation}
where $\vec n^{(m)} = \vec n - \hat e_m + \hat e_{m-1}$. For a given $M$ and $N$, we can numerically find the eigenbasis within each $N_B$ sector where only the ratio $t/U$ determines the eigenvectors. We denote the eigenvectors $|\alpha,N_B\rangle$ where each sector has a different number of allowed $\alpha$.

We now eigen-decompose the coupling operator $\mathcal B = \sum_i b_i$ within the fully-permutation-symmetric sector to find:
\begin{equation}\label{eq:Bmats}
    \mathcal B =
    \sum_\lambda \mathcal B_\lambda
    \notag = \sum_{(\alpha,\beta,N_B)}B_{(\alpha,\beta,N_B)} |\alpha,N_B-1\rangle \langle \beta N_B |.
\end{equation}
where $B_{(\alpha,\beta,N_B)}=\langle \alpha, N_B-1| \mathcal B | \beta N_B\rangle $ and $\lambda = (\alpha,\beta,N_B)$ is a combined index. 
Note that the operators in the sum have energy $[H,|\alpha,N_B-1\rangle\langle \beta N_B|] = -(E_{\beta, N_B} - E_{\alpha,N_B-1})|\alpha,N_B-1\rangle\langle \beta N_B| = -\omega_{\alpha,\beta,N_B} |\alpha,N_B-1\rangle\langle \beta N_B|$.

With the eigenoperators in hand, we can immediately write out the interaction-picture Lindblad equation (ignoring the Lamb shift)
\begin{equation}
\begin{aligned}\label{eq:Lind_no_basis}
    \dot \rho(t) &= \sum_{\lambda,\lambda' : \omega_{\lambda}=\omega_{\lambda'}} \gamma(\omega_\lambda) \mathcal L\left( B^{\vphantom{\dagger}}_{\lambda}(\omega_\lambda), B_{\lambda'}^\dagger(\omega_{\lambda'})\right)[ \rho(t)]  \\
   &+ \sum_{\lambda,\lambda' : \omega_{\lambda}=\omega_{\lambda'}} I(\omega_\lambda) \mathcal L\left( B^{\dagger}_{\lambda}(\omega_\lambda), B^{\vphantom{\dagger}}_{\lambda'}(\omega_{\lambda'}) \right)\left[ \rho(t)\right].
\end{aligned}
\end{equation}
The first (second) line corresponds to decay (pumping) processes.
Equality between $\omega_\lambda$ and $\omega_{\lambda'}$ is determined by the condition $|\omega_\lambda-\omega_\lambda'|\ll \tau^{-1}$, where $\tau$  sets the relaxation timescale of the system.
 The specifics of the environment enter only through 
\begin{equation} \label{eq:environ_corr}
\begin{aligned}
    \gamma(\omega) &= \int_{-\infty}^{\infty} ds e^{i\omega s}\langle \mathcal R(s) \mathcal R^\dagger\rangle \\
    I(\omega) &= \int_{-\infty}^{\infty} ds e^{-i\omega s}\langle \mathcal R(s)^\dagger  \mathcal R\rangle
\end{aligned}
\end{equation}
and we have assumed $\langle \mathcal R(s) \mathcal R\rangle = \langle \mathcal R^\dagger(s) \mathcal R^\dagger \rangle = 0$ (and $\tau^{-1}$ is set by $\gamma(\omega), I(\omega)$ which are assumed to be the smallest energy scales in the weak-coupling limit).

We now write out \eqref{eq:Lind_no_basis} in the eigenbasis, $|\alpha, N_B\rangle$. One difficulty in solving the resulting equation is the presence of coherences (i.e. off-diagonal terms of the density matrix in the basis of energy eigenstates). As we derive in the SI Appendix, the Lindblad master equation can only generate coherences within energy-degenerate sectors, and therefore, if there is no degeneracy, the steady-state density matrix is diagonal. However, there is a weaker condition that ensures that the $\mathcal B_\lambda(\omega_\lambda)$ operators never generate coherences when acting on a diagonal $\rho$ (see SI Appendix): $\omega_{\alpha,\beta,N_B} = \omega_{\alpha',\beta,N_B}$ ($\omega_{\alpha,\beta,N_B}=\omega_{\alpha,\beta',N_B}$) only when $B_{(\alpha,\beta,N_B)} =0$ or $B_{(\alpha',\beta,N_B)}=0$ ($B_{(\alpha,\beta,N_B)}=0$ or $B_{(\alpha,\beta',N_B)}=0$), respectively. In our analytic or numerical results, we always check that the above condition is satisfied for this system. 

We can therefore write $\rho=\sum_{\alpha, N_B}\rho_{\alpha, N_B} |\alpha, N_B\rangle \langle \alpha, N_B|$ and
\begin{equation}
\begin{aligned}\label{eq:Lind_basis}
    \dot \rho_{a,N_B} &=   \sum_{\lambda=(\alpha,a,N_B)}|B_{\lambda}|^2 \left( I(\omega_{\lambda})\rho_{\alpha,N_B-1} - \gamma(\omega_{\lambda}) \rho_{a,N_B}\right) \\
    &+ \sum_{\lambda = (a,\beta,N_B+1)} |B_{\lambda}|^2 \left( \gamma(\omega_{\lambda}) \rho_{\beta,N_B+1} - I(\omega_{\lambda}) \rho_{a,N_B}\right).
\end{aligned}
\end{equation}

As a check on our derivation, we can set $I(\omega)$ and $\gamma(\omega)$  to represent a thermal bath of free bosons. In this case, $\gamma(\omega) = A(\omega) (1+ n_B(\omega/T))$ and $I(\omega) = A(\omega) n_B(\omega/T)$ where $A(\omega)$ is a function only of $\omega$, $n_B(x) = (e^x-1)^{-1}$ is the Bose-Einstein distribution. We numerically find that the steady-state density matrix, $\dot \rho_{a, N_B} = 0$ has the thermal form $\rho_{a,N_B} \sim e^{-E_{a,N_B}/T}$. In general, we consider non-zero $I(\omega)$ as resulting from light-induced, rather than thermal, production of excitons.

\section{Key observables, phases, and exponents}

Once the steady state satisfying $\dot \rho_{a,N_B} = 0$ from \eqref{eq:Lind_basis} is determined, there are two key observables to evaluate: 
\begin{equation}
    n_B = \frac{1}{N}\sum_i \langle b_i^\dagger b_i\rangle,\qquad \Psi^2 = \frac{1}{N^2} \sum_{i,j} \langle b_i^\dagger b_j\rangle = \frac{\langle \mathcal B^\dagger \mathcal B\rangle}{N^2}. 
\end{equation}
The two phases present in our model are Mott-insulating phases with integer exciton density $n_B\in \{0,1,2,...\}$ (and $\Psi^2=0$) and a superfluid phase with generic $n_B$ and $\Psi^2\ne 0$.
 In the equilibrium case, the superfluid phase is determined by the non-zero expectation $\langle b_i \rangle$ (or equivalently the long-range order of $\langle b_j^\dagger b_i\rangle$), and $\Psi^2$ is our analogous observable.

In addition to the phases themselves, we study the phase transitions and extract the critical exponents. We will tune across the phase transitions primarily by adjusting the ratio $r(\omega)=I(\omega)/\gamma(\omega)$ of the exciton production and decay strengths. With $\mu \gg U,t$, we may Taylor expand $r(\omega)$ around $\omega=\mu$ and the transition occurs when $r(\mu)=r_1$ reaches a critical value $r_c$.

At a continuous phase transition, there is an order parameter $\chi$ that takes a particular form close to the phase transition
\begin{equation}
    \chi  =  \begin{cases} \chi_0(r_1-r_c)^\beta & r_1 > r_c \\
    0 & r_1 < r_c
    \end{cases},
\end{equation}
where $\chi_0$ is a constant, and the correlation length $\xi$ of the system will diverge as $\xi \sim (r_1-r_c)^{-\nu}$. The exponents $\beta$ and $\nu$ are universal numbers set by the phase transition itself, independent of the exact microscopic model. For systems of a finite length $L$, this diverging correlation length leads to the finite-size scaling form for $\chi = L^{-\beta/\nu} f_{\chi}(L^{1/\nu}(r_1-r_c))$, where $f_\chi(x)$ is a finite-size scaling function \cite{Cardy}. 

The all-to-all hopping in our system makes it difficult to define a correlation length, so we define the exponent $\lambda$ as the analog for $\nu$ in our system. At the Mott-Mott transition from the $n_B=0$ to $n_B=1$ Mott phases, the order parameter is simply $n_B$, which therefore has the form
\begin{equation}\label{eq:nbscalefunc}
    n_B = N^{-\beta_{m}/\lambda_m} f_{n_B,m}(N^{1/\lambda_m} (r_1-r_c) )
\end{equation}
where $\beta_m$, $\lambda_m$ are the critical exponents and $f_{n_B,m}(x)$ is the scaling function for $n_B$ at the Mott-Mott transition. Similarly, at the Mott-superfluid (Mott-Mott) transition, the order parameter is $\Psi^2$ and takes the scaling form 
\begin{equation}
    \Psi^2 = N^{-\beta_s/\lambda_s}f_{\Psi^2,s}(N^{1/\lambda_s}(r_1-r_c))
\end{equation}  with similarly defined $\beta_s$, $\lambda_s$ and $f_{\Psi^2,s}(x)$. Note that the exponents $\lambda_{m/s}$ and $\beta_{m/s}$ and scaling functions do not depend on the exact definition of the tuning parameter used to move between the two phases, and they can distinguish between different critical points. 

\section{Analytic results}
Due to needing to find the eigenvectors of $H_0$, \eqref{eq:H0eigprob}, we cannot make additional analytic progress in most cases. We will treat the general problem numerically in the next section, but there are two cases where more analytic progress can be made \footnote{We additionally consider $U=0$ in the SI Appendix but our formalism must be slightly modified to admit a consistent calculation.}.

\subsection{$t=0$}

 In this limit, the eigenvectors of $H_0$ are simply the $|\vec n\rangle$ states defined above. Instead of simplifying \eqref{eq:Lind_basis}, we note that
\begin{equation}\label{eq:B_in_Os}
    \mathcal B = \sum_{j=1}^M \sqrt{j} \mathcal O_{j-1,j}
\end{equation}
and $[H,\mathcal O_{j-1,j}] = -[\mu + U(j-1)]O_{j-1,j}=-\omega_jO_{j-1,j}$. We can therefore write out the equivalent of \eqref{eq:Lind_no_basis}
\begin{equation}
\begin{aligned}
    \dot \rho(t) &= \sum_j \gamma(\omega_j) j \mathcal L\left(\mathcal O_{j-1,j},\mathcal O_{j-1,j}^\dagger\right)[\rho(t)] \\
    &+\sum_j I(\omega_j) j \mathcal L\left(\mathcal O_{j-1,j}^\dagger,\mathcal O_{j-1,j}\right)[\rho(t)].
\end{aligned}
\end{equation}

We now substitute $\rho = \sum_{\vec n,\vec m} \rho_{\vec n, \vec m} |\vec n\rangle \langle \vec m|$.
Using \eqref{eq:O_act_n} and the fact that we start from a vacuum state, one can check that
the condition $\rho_{\vec n,\vec m}=0$ when $\vec n \ne \vec m$, is preserved under time evolution. Therefore, we may instead write $\rho(t)=\sum_{\vec{n}}\rho_{\vec{n}}(t)|\vec n\rangle \langle \vec n|$. 

We can explicitly write out the steady-state equation as
\begin{equation}
\begin{aligned}
    0&=\sum_{j=1}^M j n_{j-1} (n_j+1) ( \gamma(\omega_j)\rho_{\vec n-\hat e_{j-1} + \hat e_j}-I(\omega_j) \rho_{\vec n})\\
    &+\sum_{j=1}^M j n_j (n_{j-1}+1) (I(\omega_j) \rho_{\vec n - \hat e_j + \hat e_{j-1}} - \gamma(\omega_j) \rho_{\vec n})
\end{aligned}
\end{equation}
where the above equation holds for all $\vec n$ and  $n_j$ denote components of $\vec n$. A sufficient ``detailed-balance''-like condition for the steady-state is that all terms in the sum are zero,
\begin{equation}\label{eq:all_terms_zero}
   \gamma(\omega_j) \rho_{\vec n - \hat e_{j-1}+ \hat e_j} =I(\omega_j) \rho_{\vec n} ,
\end{equation}
which remarkably specifies the solution in this case. (In general, the equivalent set of equations will not hold as they overdetermine the solution.) It can easily be checked that the steady-state is given explicitly as
\begin{equation}\label{eq:ss_teq0}
\begin{aligned}
    \rho_{\vec n} &= \rho_{0} \prod_{i=1}^M\left(\prod_{j=1}^i\frac{I(\omega_j)}{\gamma(\omega_j)}\right)^{n_i} =\rho_0\prod_{i=1}^M\left(r(\omega_i)\right)^{\sum_{j=i}^M n_j}
\end{aligned}
\end{equation}
where we have defined $r(\omega_j)\equiv I(\omega_j)/\gamma(\omega_j)=r_j $.
This form of the steady-state manifestly reveals that we can set the maximum number of bosons per site we need to consider, $M$, by the condition $r(\omega_M) \ll 1$, when $t=0$. 

We now analyze the case $M=2$ in detail: We can rewrite the density matrix as $\rho_{\vec n} = \rho_0 r_1^{n_1} (r_1r_2)^{n_1+n_2}$ (and we assume that $r_j>r_{j+1}$). We can compute the average number of bosons per site $n_B = \langle N_B\rangle/N$ as
\begin{equation}
\begin{aligned}
  n_B&= \begin{cases}
       0 +\frac{1}{N}\frac{r_1+(2-3r_1)r_1r_2}{(r_1-1)(r_1r_2-1)} + ... & \text{if $r_1^N,r_2^N \ll 1$}\\ 
      1+ \frac{1}{N}\frac{1-r_1 r_2}{(r_1-1)(r_2-1)}+ ... & \text{if $r_1^{-N},r_2^N \ll 1$} \\
       2+\frac{1}{N}\frac{3-2r_2 -r_1r_2}{(r_2-1)(r_1r_2-1)} + ... & \text{if $r_1^N, r_2^N \gg 1$}.
   \end{cases}
\end{aligned}
\end{equation}
We can see, in the thermodynamic limit, that our model has a sharp transition from an $n_B=0$ to $n_B=1$ to $n_B=2$ state as we increase the ratio $r_1$ and $r_2$. We identify these regions as Mott-insulators (and we can numerically check that $\Psi^2\to 0$). We can analyze these critical points by evaluating
\begin{equation}
    n_B = \begin{cases} \frac{1}{2}  +\frac{1}{12}(r_1-1)N +... & \text{if $r_1 \approx 1$, $r_2^N \ll 1$} \\
    \frac{3}{2} + \frac{1}{12}(r_2-1)N + ... & \text{if $r_2 \approx 1$, $r_1^N \gg 1$}
    \end{cases}
\end{equation}
implying that $n_B$ has the scaling form $n_B=N^{-\beta_m/\lambda_m}f_{n_B,m}(N^{1/\lambda_m}(r_j-1))$  with $\lambda_m=1$ and $\beta_m=0$.  We expect that the values of $\lambda_m,\beta_m$ are insensitive to increasing $M$.

In equilibrium, this transition is first order, since it is simply a level crossing of the two competing ground states, $|N \hat e_j\rangle$ to $|N\hat e_{j+1}\rangle$. However, in the nonequilibrium setting, as we approach the phase transition there are more and more states in the Hilbert space that contribute to the steady-state density matrix, which allows for a  continuous phase transition to emerge between $\rho_{ss}\approx |N\hat e_j\rangle \langle N \hat e_j|$ to $\rho_{ss}\approx |N\hat e_{j+1}\rangle \langle N \hat e_{j+1}|$.

\subsection{$M=1$}

If we consider the case of hardcore bosons, i.e. $U\to \infty$, we can make additional progress as well because there is only one state per (fully symmetrized) sector, $N_B$, and therefore the eigenstates of $H_0$ can be specified as $|N_B\rangle$. The eigenenergy of the states is given by
\begin{equation}
    E_{N_B}= \mu N_B - \frac{t}{N} N_B (N- N_B + 1)
\end{equation}
and we can evaluate
\begin{equation}
    \mathcal B = \sum_{N_B=1}^N \sqrt{N_B(N-N_B+1)}|N_B-1\rangle \langle N_B| 
\end{equation}
so $B_{\lambda}=B_{N_B} = \sqrt{N_B(N-N_B+1)}$
and these have energies $\omega_{N_B} = E_{N_B} - E_{N_B-1} = \mu - t(1-2(N_B-1)/N)$. When $t$ is non-zero, we see that the energy will take on a range of values from $\mu -|t| < \omega < \mu + |t|$ in the thermodynamic limit.

Since all $\omega_{N_B}$ are unique, we still have no coherences and can write down the density matrix equation
\begin{equation}
\begin{aligned}
    \dot \rho_m &= m (N-m+1)( I(\omega_{m}) \rho_{m-1} - \gamma(\omega_m) \rho_m) \\
    &+ (m+1) (N-m) (\gamma(\omega_{m+1}) \rho_{m+1} - I(\omega_{m+1}) \rho_m)
\end{aligned}
\end{equation}
where $m$ is specifying $N_B$. Again noting that a sufficient condition for the steady-state solution is
\begin{equation}\label{eq:all_terms_zero_M1}
    \gamma(\omega_m)\rho_m = I(\omega_m) \rho_{m-1}
\end{equation}
we find
\begin{equation}\label{eq:rhoss_m1}
\rho_m = \rho_0\prod_{j=1}^m \frac{I(\omega_j)}{\gamma(\omega_j)} =
\rho_0 \prod_{j=1}^m r(\omega_j) 
\end{equation}
If $t=0$, we see that $\omega_m = \mu$ for all $m$, so we quickly derive \eqref{eq:ss_teq0} for $M=1$.

We will now show that, in addition to the $n_B=0$ and $n_B=1$ Mott phases seen before, there is an intermediate phase. 
We first assume the pump-to-decay ratio $r(\omega_j)=r_j$ decreases monotonically with $\omega$, which fits the intuition that higher energy excitations are prone to decay more quickly. 
Defining $r_<\equiv r(\mu-t)$ [$r_>\equiv r(\mu+t)$],  we can recover the Mott phases from before by computing
$\rho_m =\rho_0 r_<^m e^{ (r_<' t m^2)/(r_< N)}$ $[\rho_{N-m} =\rho_N r_>^{-m} e^{ (r_>' t m^2)/(r_> N)}]$ when $r_< \ll 1$ ($r_> \gg 1$), respectively,  
where $r_{</>}'$ are the corresponding values of $r'(\omega)=dr(\omega)/d\omega <0$ (see SI Appendix for additional calculation details for this section).

Next, we consider the case $r_< > 1 > r_>$.
There then exists some index $j^*$ where 
$r_{j^*} = 1$.
From \eqref{eq:all_terms_zero_M1} and  \eqref{eq:rhoss_m1}, we can find as $N\to \infty$ 
\begin{equation} \label{rho:sf}
    \rho_{j^*\pm m} \approx \rho_{j^*} e^{-\delta m^2}
\end{equation}
where $\delta = -r'_{j^*} t/N > 0$ from the monotonic decrease assumption.

We can deduce this density matrix represents an additional phase by measuring $\Psi^2$. Since $M=1$, we see that $\mathcal B^\dagger \mathcal B|N_B\rangle = N_B (N-N_B+1)|N_B\rangle$. In the three aforementioned limits, we find 

\begin{equation}
    \Psi^2 = \begin{cases}
    \frac{r_<}{N}(1-r_<)^{-1}+ ... & \text{if $r_< \ll 1$} \\
    \frac{j^*}{N}\left(1-\frac{j^*}{N}\right)+ ... 
    & \text{if $r_{j^*}=1$ and $j^*\sim O(N)$} \\
    \frac{1}{N}(1-r_>^{-1})^{-1} + ...& \text{if $r_> \gg 1$} 
    \end{cases}
\end{equation}

As above, we can observe a critical scaling emerge as $r_<\to 1$. Using \eqref{eq:rhoss_m1}, we can find 
\begin{equation}
    \Psi^2 
    \approx \frac{1}{N^{1/2}}
    \frac{\int_0^{\infty}dx x \exp{\left[\frac{r_<'t}{r_<}x^2-(1-r_<)\sqrt{N}x\right]}}{\int_0^{\infty}dx \exp{\left[\frac{r_<'t}{r_<}x^2-(1-r_<)\sqrt{N}x\right]}}.
\end{equation}
which has the scaling form
$\Psi^2 =N^{-1/2}f_{\Psi^2,s}(N^{1/2} (r_<-1))$, 
implying that $\lambda_s=2$ and $\beta_s=1$. A similar finite-size scaling form, with the same value of $\lambda_s$, can be derived for $n_B$.

\section{Numerical Results}
\label{sec:numer}

From the analytic results, we can see there are two phases distinguished by  $\Psi^2$. If $\Psi^2\to 0$, then we are in a Mott insulating phase where $n_B$ is approximately an integer. Otherwise, $n_B$ takes an intermediate value and $\Psi^2$ is non-zero, like in the superfluid phase. By solving the system numerically, we can explore what happens when $t\ne0$ and more than one boson per site is allowed.

In order to proceed numerically, we must specify $I(\omega)$ and $\gamma(\omega)$. We take $\gamma(\omega) = \gamma_0 (\omega/\mu)^3$, the correct form for photon-driven decay, and we assume $I(\omega)=I_0$  for simplicity \footnote{The realistic $I(\omega)$ would require sophisticated modeling since the excitons are generated through complicated pump and relaxation processes.}. For the moir\'e TMD system \cite{xiong2023correlated}, the realistic excitation gap $\mu\approx 1.5$ eV and the repulsion $U\approx 30$ meV. However, we fix $\mu=150$ meV as $\mu$ only affects $\gamma'(\omega)$ [and consequently $r'(\omega)$] which we need to be large to achieve numerical convergence. We let $(t+U)$ set the energy scale, and vary both $t/(t+U)$ and $r(\mu)=I_0/\gamma_0$ to map out the phase diagram and study the critical exponents.

Once $t\ne 0$ and $M\ge 2$, we can only numerically determine the Lindblad operators and energy eigenstates, and the equivalent conditions to \eqref{eq:all_terms_zero} and \eqref{eq:all_terms_zero_M1} cannot all be satisfied. Instead, we construct the Liouvillian superoperator $\mathcal L$ satisfying $\dot \rho = \mathcal L \rho$ from \eqref{eq:Lind_basis} and search for the steady-state $\rho$ that has eigenvalue zero. We can  numerically evaluate the $M=2$ case up to large system size of $N\sim 1000$. In this case, when $t=0$, the various $N_B$ sectors are non-degenerate implying that the steady-state density matrix is diagonal, as we argued above. We check numerically that this non-degeneracy persists when $t\ne 0$.

 In addition to measuring $n_B$ and $\Psi^2$, we measure the entropy $S=-\text{Tr}[\rho \ln(\rho)]$ and the Liouvillian gap, $\Delta_{\mathcal L}$ determined from the smallest real part of any of the non-zero eigenvalues of $\mathcal L$ \cite{Zhang2022}. These two quantities allow us to better distinguish between the two phases and characterize their critical behavior numerically. We extrapolate these quantities to the $N\to \infty$ limit (see SI Appendix).

In Fig.~\ref{fig:pd_main}, we observe the same two phase as above: the Mott phase is signified by integer $n_B$ with $\Psi^2 =0$ and $S/N = 0$,
and the superfluid phase is signified by generic $n_B$ and $\Psi^2\ne 0$. At the phase boundary, $S/N$ attains a maximum, clearly demarcating the two phases.  The phase diagram in Fig.~\ref{fig:pd_main} qualitatively matches the equilibrium phase diagram with its $n_B=1$ Mott lobe. If $M>2$, then we would see additional Mott lobes as well (see SI Appendix). 

\begin{figure}[htb]
    \includegraphics[width=0.9\linewidth]{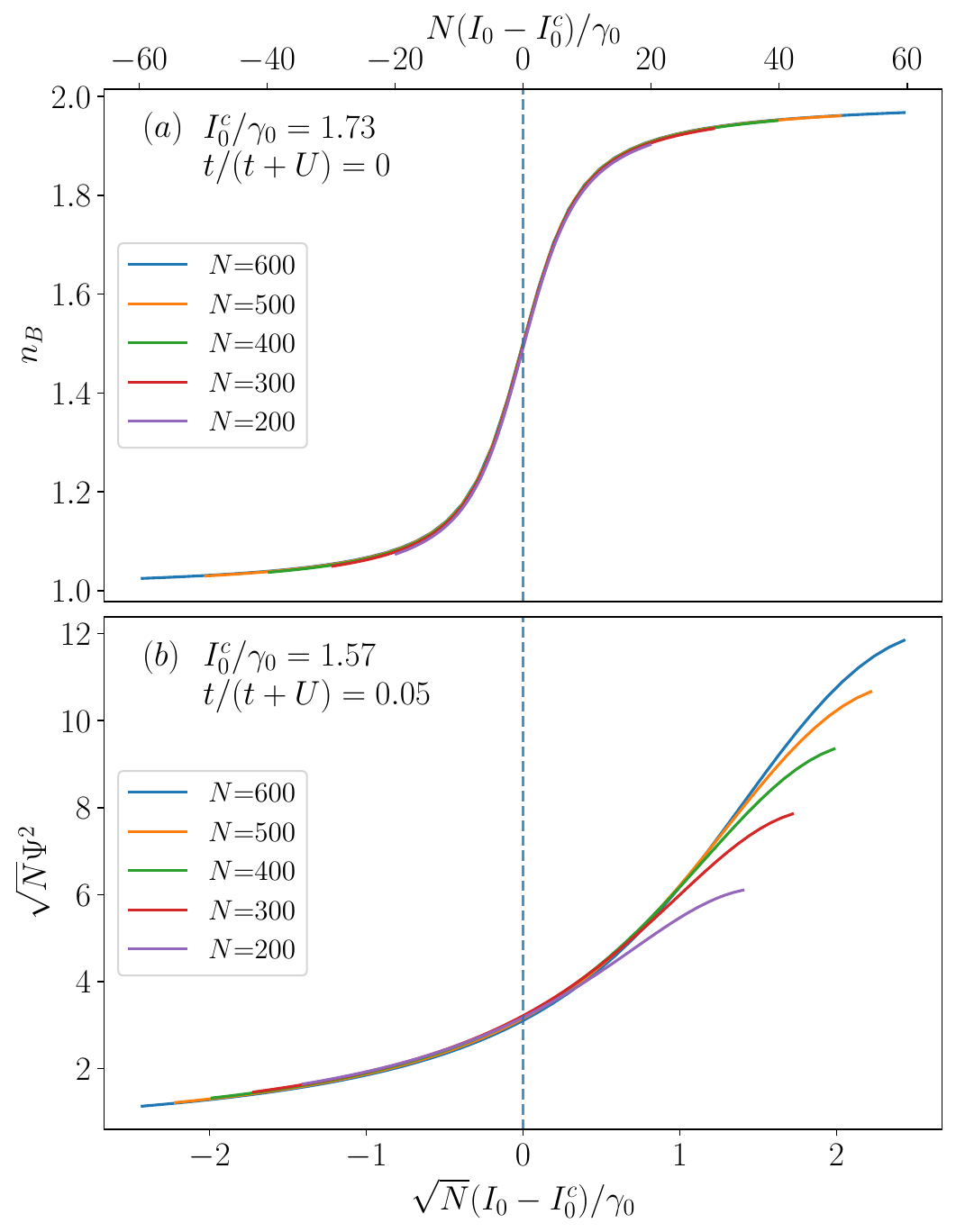}\\
    \caption{We perform scaling collapses for the order parameters at the (a) transition between the $n_B=1$ and $n_B=2$ Mott phases and (b) transition between Mott and supefluid phases. The $N\to \infty$ curve that the different data collapses onto in (a) [(b)], are, respectively, the $f_{n_B,m}(x)$ [$f_{\Psi^2,s}(x)$] defined in the text.  We find that $\lambda_m=1$ and $\beta_m=0$  for the former and $\lambda_s=2=\beta_s+1$ for the latter. We have confirmed that this exponent stays the same regardless of which Mott lobes are involved and regardless of whether we tune $I_0/\gamma_0$ or $t/U$.}
    \label{fig:scaling_col}
\end{figure}

In addition to mapping out the phase diagram, we can confirm the universal scaling forms discussed above. In Fig.~\ref{fig:scaling_col}(a), at the transition between two Mott phases, we observe that $n_B = f_{n_B,m}(N(I_0-I_0^c))$, as expected from \eqref{eq:nbscalefunc}, indicating that $\lambda_m=\beta_m+1=1$. 
Within the equilibrium phase diagram, this transition is first order (a level crossing)  indicating that the environment can change the nature of the critical point. 
In Fig.~\ref{fig:scaling_col}(b), we similarly see $\Psi^2 = N^{-1/2}f_{s}(N^{1/2}(I_0-I_0^{c}))$, as above, for a transition between the Mott and superfluid phase with $\lambda_s=2$ and $\beta_s=1$, as opposed to the equilibrium values of $\lambda_s=\beta_s=1$ (see SI Appendix).
We confirm that the same exponents are extracted while tuning $t/U$ instead of $I_0/\gamma_0$ across the phase transition.

Finally, when $M=1$, we can extract the Liouvillian gap, $\Delta_{\mathcal L}$. Away from the critical point, $\Delta \mathcal L/N$ approaches a constant indicating an effect similar to superradiance \cite{breuer2002theory}. 
At criticality, however, we find $\Delta \mathcal L/N \sim N^{-1/\lambda}$, indicating a critical slowing down (i.e. a longer time to reach the steady state) near the phase boundary (see Fig.~S7 in the SI Appendix).
We expect that this same behavior should arise at the phase transitions when $M\ge 2$, but we are unable to compute $\Delta \mathcal L$ for large enough system sizes.

\section{Discussion and Conclusion}

We have introduced a solvable variant of the Bose Hubbard model coupled to an environment. We find two phases that resemble the equilibrium Mott and superfluid phases, but the density matrix does not take a Boltzmann form [\eqref{rho:sf}]. We extract critical exponents $\lambda_m=\beta_m+1=1$ ($\lambda_s=\beta_s+1=2$) for the transition between a Mott phase and a Mott (superfluid) phase, respectively. These exponents are different from the observed equilibrium exponents  derived in the SI Appendix (and the Mott-Mott transition is first-order in equilibrium). The change in criticality is not surprising--although previous work found that the open system criticality can only change a dynamic exponent \cite{Sieberer2013,Sieberer2014}, universality is typically only insensitive to local perturbations, and the environment acts as a global perturbation.

We derive the Lindblad operators and master equations from an explicit form of the coupling between the system and environment. Our model is solvable because the environment couples in a site-invariant way that allows us to only consider the fully sysmmetric sector of the permutation group, and the calculation is controlled by $1/N$ for large system size.  We note that, although the full eigensystem information of $H_S$ is difficult to obtain, only the eigenstates that are connected by the derived Lindblad operators are relevant, and we demonstrate this idea in the SM for the case $U/t=0$ in the SI Appendix. For generic $t/U$, it may be possible to obtain Lindblad operators perturbatively controlled by $1/N$, but we leave this work for the future.

Our formulation of site-invariant coupling between the system and its environment is reminiscent of the Dicke model \cite{kirton2019introduction} and will display superradiance \cite{breuer2002theory}. However, since we model the production as the reverse process of decay, our model also has superabsorption. This observation explains why, when $I<\gamma$, the steady state has only a finite number of excitons in the thermodynamic limit: if the production rate $I$ is less than the decay rate $\gamma$, an extensive number of excitons cannot be built up. 

In the experimental systems \cite{park2023dipole,lian2024valley,gao2024excitonic,xiong2024tunable,xiong2023correlated}, there may be an additional local coupling $H_I' = \sum_i b_i^\dagger \mathcal R_i' + \hc$ responsible for the production of the bosons, but introducing such a term into our formalism would destroy the solvability. Future work should consider the competition between a local production of  bosons and their global (or longer-range) decay, as well as a short-range hopping system Hamiltonian, which might be resolved by numerical methods such as dynamical mean-field theory.

\begin{acknowledgements}
We thank Chenhao Jin and Richen Xiong for discussing their experiments. We thank Matthew Fisher for helpful discussions.  
LB and ZS are supported by the NSF CMMT program under Grant No. DMR-2419871.  TC is supported by a University of California Presidential Postdoctoral Fellowship and acknowledges support from the Gordon and Betty Moore Foundation through Grant No. GBMF8690 to UCSB. This research was supported by the Simons Collaboration on Ultra-Quantum Matter, which is a grant from the Simons Foundation (Grant No. 651440), and, in part, by grant NSF PHY-2309135 to the Kavli Institute for Theoretical Physics (KITP). Use was made of the computational facilities administered by the Center for Scientific Computing at the CNSI and MRL (an NSF MRSEC; DMR-1720256) and purchased through NSF CNS-1725797. 
\end{acknowledgements}

\bibliography{Citations.bib}

\appendix
\setcounter{figure}{0}
\renewcommand\thefigure{S.\arabic{figure}}

\section{Condition for diagonal steady state}
Given a generic Hamiltonian $H=H_S + H_R+\sum_\alpha S_\alpha \mathcal \otimes R_\alpha$, with $S_\alpha$ a system operator and $\mathcal R_\alpha$ a reservoir operator, we can separate it into a system ($H_S$), reservoir ($H_R$), and interaction piece. The resulting master equation for the density matrix of the system has the form in the weak-coupling limit \cite{breuer2002theory}
\begin{equation}
    \dot \rho_S =-i[H_S+H_{LS},\rho_S] +\sum_{\omega,\alpha,\beta} \gamma_{\alpha,\beta}(\omega) \mathcal L\left(S_{\beta}(\omega), S_{\alpha}^\dagger(\omega)\right)[\rho]
\end{equation}
where $S_{\alpha}(\omega)=\sum_{E'-E=\omega} \Pi(E)S_\alpha \Pi(E')$ with $\Pi(E)$ the projector on the energy $E$ eigenspace of $H_S$, and $H_{LS}=\sum_E \Pi(E) H_{LS,E} \Pi(E)$ is the Lamb shift Hamiltonian. As before $\gamma_{\alpha,\beta}(\omega)$ is expressed as correlators of the $R_{\alpha,\beta}$.

We now will assume that the density matrix only has coherences within each eigensector and show that it remains in this form under time evolution. To do so, we write the density matrix as $\rho_S = \sum_E \Pi(E) \rho(E) \Pi(E)$ where $\rho(E)$ is density matrix acting on the energy $E$ eigenspace. For this form, we can explicitly compute
\begin{equation}
\begin{aligned}
    \dot \rho_S&=-i[H_S + H_{LS},\rho_S]+\mathcal D[\rho]
    \notag\\
    &=\sum_E \Pi(E) [H_{LS,E},\rho(E)]\Pi(E) + \mathcal D[\rho]
\end{aligned}
\end{equation}
with dissipative part

\begin{widetext}
\begin{equation}
\begin{aligned}
 \mathcal D[\rho] = \sum_{\omega,\alpha,\beta}\gamma_{\alpha,\beta}(\omega) 
 \sum_{E'-E=\omega}&
 \bigg( \Pi(E) S_\beta\Pi(E') \rho(E') \Pi(E') S_\alpha^\dagger \Pi(E)  \\
    -&\frac{1}{2} \Pi(E') S_{\alpha}^\dagger \Pi(E) S_\beta \Pi(E') \rho(E') \Pi(E')
    -\frac{1}{2} \Pi(E')\rho(E') \Pi(E') S_\alpha^\dagger \Pi(E) S_\beta\Pi(E')\bigg)
\end{aligned}
\end{equation}
\end{widetext}
We therefore see that the density matrix remains block diagonal in the different energy sectors. We can always search for a steady state of this form, especially since we can always start from a ground state.
We then turn on incoherent pumping, which will never create coherences between different energy blocks. The only way to create and maintain such coherences in this limit would be through a time-dependent $H_S$.

An immediate consequence of this result is that if the spectrum of $H_S$ is non-degenerate, then the steady-state density matrix will be diagonal, $\rho = \sum_E \rho_{E} |E\rangle \langle E|$. However, this condition is not necessary. 

In our above analysis, we always have diagonal damping rates $\gamma_{\alpha,\beta}(\omega)=\gamma_{\alpha,\alpha}(\omega)=\gamma_\alpha(\omega)$ with $\gamma_{+}(\omega) = I(\omega)$ [$\gamma_-(\omega)=\gamma(\omega)]$ for corresponding operator $S_+ = \mathcal B^\dagger$ ($S_-=\mathcal B$), respectively. 
To find a diagonal $\rho$ in this case, we just need that in some basis of each eigensector, $S_{\pm}(\omega)|E,m\rangle \propto |E\pm \omega,k\rangle$ (where $m$, $k$ are indexing the degeneracy), 
i.e.~the choice of basis leads to jump operators that do not create coherences between states of the same energy.
This condition ensures that a density matrix $\rho = \sum_{E,m} \rho_{E,m}|E,m\rangle\langle E,m|$ stays of this form. Notably, since there is a conserved quantity $N_B$ in our model that is changed by the $S_\pm$, it is clear that this condition will always hold if there is no degeneracy within each fixed $N_B$ sector. If there is such degeneracy, the density matrix will still have a diagonal form if $S_\pm(\omega)$ does not generate superpositions of degenerate states. Writing this condition in the notation of the main text gives the result as stated there.

\section{Standard assumption of Lindblad operators}

Let's analyze the case with $t=0$ with the form of the jump operators from e.g. Refs~\cite{LeBoite2013,LeBoite2014}. We have
\begin{equation}
\begin{aligned}
H_0 &= \mu \sum_i \bigl(n_i + \frac{U}{2}n_i (n_i-1)\bigr) \\
    \dot \rho(t) &= -i[H_0,\rho] + \sum_i 
    \bigl( \gamma \mathcal L(b_i, b_i^\dagger)[\rho] + I\mathcal L(b_i^\dagger, b_i)[\rho] \bigr)
    \end{aligned}
\end{equation}
In this case, the sites are decoupled and $\rho = \bigotimes_i \rho_i$. We can therefore consider only one site at a time. We write out the density matrix in the number basis $\rho_i = \sum_{n=0}^M\rho_{i,n}|n_i\rangle \langle n_i|$, whose form is preserved under time evolution,  and we find
\begin{equation}
\dot \rho_{i,n} = (n+1) (\gamma \rho_{i,n+1} - I \rho_{i,n}) + n(I\rho_{i,n-1} - \gamma \rho_{i,n}).
\end{equation}
As in the main text, we see that the form $\rho_{i,n} = \rho_{i,0}(I/\gamma)^n$ is a solution. 

There are at least two shortcomings to this formalism, however, that are apparent even in this simple example: first, this steady state appears to not depend at all on the parameters of $H_0$ (as opposed to our solution where $U/\mu$ affects the location of the transitions). Second, it is not clear how to include temperature. If we assume that the bath has a temperature $T$, there is no clear way to include it in the definition of $I$ and $\gamma$ because they lack a dependence on the energy. One natural way would be to assume $I/\gamma = e^{-\mu /T}$, but the fact that the energy is quadratic in $n$ means that the steady state cannot fit the thermal form $\rho_n = e^{-(\mu n+ U n(n-1)/2)/T}$. As we discuss in the main text, our approach readily capture the coupling to a thermal reservoir, which leads to the steady-state density matrix being of the Boltzmann form $\rho \sim e^{-H/T}$, a well-known attribute of the weak-coupling limit Lindblad formalism \cite{breuer2002theory}.

\section{Short-ranged or long-ranged Lindblad operators?}

In our text, we assumed that the wavelength of light was much larger than the extent of the system and therefore we took a translation-invariant coupling to our sites. In most other work, e.g. \cite{LeBoite2013,LeBoite2014}, the opposite limit is chosen. How do we know which limit is applicable and/or how do we extrapolate between them?

The derivation of the master equation provides the answer. Let's assume the environment is solely composed of photons and   the system and environment couple in the natural way, as before \cite{Hanai2018}
\begin{equation}
    H_I = \sum_{\pmb k,\pm} g_{\pmb k,\pm} a_{\pmb k,\pm} b_{\pmb k}^\dagger + \text{H.c.} = \sum_{i} \mathcal R_{i} b_{i}^\dagger + \text{H.c.}
\end{equation}
where $\mathcal R_i = \frac{1}{\sqrt{N}}\sum_{\pmb k,\pm} g_{\pmb k,\pm} a_{\pmb k,\pm} e^{i\pmb x_i \cdot \pmb k}$.

Regardless of what the system Hamiltonian is (assuming it satisfies the assumptions needed for the derivation of the master equation), the environment will only enter through the computation of
\begin{equation}
\begin{aligned}
    \gamma_{i,j}(\omega) &= \int_{-\infty}^\infty ds e^{i\omega s}\langle \mathcal R_{i}(s) \mathcal R^\dagger_{j}\rangle;
    \notag \\
    \qquad I_{i,j}(\omega) &= \int_{-\infty}^\infty ds e^{i\omega s}\langle \mathcal R_{i}^\dagger(s) \mathcal R_{j}\rangle
\end{aligned}
\end{equation}
The computation is almost identical, so we just focus on 
\begin{equation}
    \gamma_{i,j}(\omega) = \frac{1}{N} \sum_{\pmb k,\pm} 2\pi \delta(\omega-k) |g_{\pmb k,\pm}|^2 e^{i(\pmb x_i - \pmb x_j) \cdot \pmb k} n_{k}
\end{equation}
where we assumed that $H_R = \sum_{\pmb k,\pm} k a_{\pmb k,\pm}^\dagger a_{\pmb k,\pm}$ and a thermal distribution of light with $\langle a_{\pmb k,\pm}^\dagger a_{\pmb k,\pm}\rangle = n_{k}$. In order to evaluate further, we need to know what the $g_{\pmb k,\pm}$ are, which depend on the details of the system. 
Here we just assume it varies smoothly with $\pmb k$.
We note that, when $|\pmb x_i - \pmb x_j|k\gg 1$, with $k$ set by $\omega$, the angular integral will lead to terms that average to zero for $i\neq j$, and thus the dissipative part takes the form $\sum_i \gamma \mathcal{L}(L_i,L_i^{\dagger})$, i.e.~local pump and decay apply here.
However, when $|\pmb x_i - \pmb x_j|k\ll 1$, it is clear that the jump operators arising from $\mathcal R_i$ and $\mathcal R_j$ will mix together, and then we have a dissipative term of the form $\gamma \mathcal L(\sum_i L_i^{\vphantom{\dagger}},\sum_j L_j^{\dagger})$. In the latter case, we just recover the situation considered in the main text, i.e.~a global pump and decay. 

\section{Analytic solution of $U/t=0$ limit}
\begin{figure}
\centering 
\includegraphics[width=0.45\textwidth]{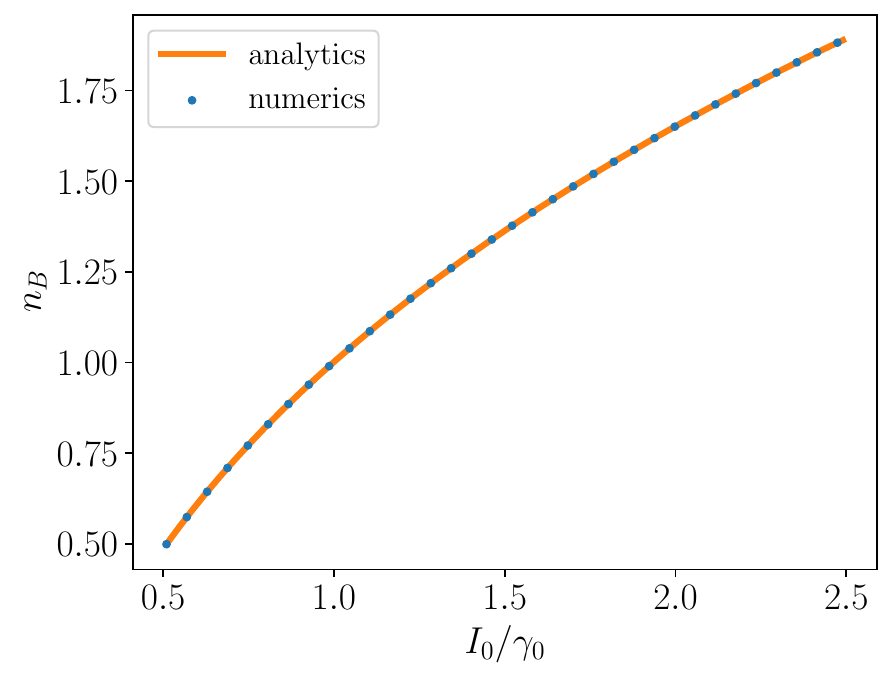}
    \caption{The analytical and numerical results for $U=0$.}
    \label{fig:U=0}
\end{figure}
As we stated in the text, the full eigensystem information of $H_S$ can be hard to obtain, but only the eigenstates that are connected by the derived Lindblad operators are relevant. Here we demonstrate this idea in the $U/t=0$ limit of the Bose Hubbard model, where we can also obtain an anlytical solution of the steady state.
In this case, it's better to reformulate the system Hamiltonian as a spin-$M/2$ system, so that we can have a direct comparison with the numerics which imposes a boson number truncation $M$.
To do so, we just need to redefine the action of $b_i$ in $H_0$ by
\begin{equation}
\begin{aligned}
    b_i |n_i\rangle &= \sqrt{n_i(M+1-n_i)} |(n-1)_i\rangle;
    \notag \\
    \qquad b_i^\dagger |n_i \rangle &= \sqrt{(n+1)(M-n)} |(n+1)_i\rangle;
    \notag \\
    \qquad \hat n_i |n_i\rangle &= n_i |n_i \rangle.
\end{aligned}
\end{equation}
These map to spin-$M/2$ operators $b_i^\dagger \to S_i^+$, $b_i \to S_i^-$, and $\hat n_i \to  S^z_i+M/2$ satisfying the usual commutation relations.
We then have the system Hamiltonian
\begin{equation}
    H_s=\mu \sum_i \hat n_i -\frac{t}{N}\sum_{i,j}b_i^{\dagger}b_j^{\vphantom{\dagger}}
\end{equation}
We still define the coupling operator $\mathcal B =\sum_i b_i$. One can then verify the commutation relation:
\begin{equation}
    [H_s,\mathcal B]=(-\mu+Mt-\frac{2t}{N}\hat{N}_B) \mathcal B.
\end{equation}
Note that $\mathcal B$ is not an eigenoperator since $\hat{N}_B$ is an operator. 
But we can decompose $\mathcal B$ as
\begin{equation}
    \mathcal B=\sum_{N_B}\mathcal B_{N_B}
\end{equation}
where $\mathcal B_{N_B}\equiv P_{N_B-1}\mathcal B P_{N_B}$, and $P_{N_B}$ is the projector onto the sector of boson number $N_B$. Then, one can verify \begin{equation}
    [H_s,\mathcal B_{N_B}]=(-\mu+Mt-2t\frac{N_B-1}{N})\mathcal B_{N_B}
\end{equation}
and thus $\mathcal B_{N_B}$ are eigenoperators, with energies 
$\omega_{N_B}=\mu-t(M-\frac{2(N_B-1)}{N})$,
which have the same form as in the $M=1(U/t\neq0)$ case in the text.
In the $M=1(U/t\neq0)$ case, each $N_B$ sector has only one eigenstate and thus we can label them by $|N_B\rangle$ and write down the master equation in such basis. 
Here in each $N_B$ sector, there are clearly more than one eigenstate. However, since we start from the vacuum state, all the states visited during the Lindblad time evolution are those generated by the Lindblad operators, i.e.~we may still denote $|N_B\rangle \equiv C_{N_B} (\mathcal B)^{N_B} |\text{vac}\rangle$, with $C_{N_B}$ being the normalization factor. 
One can check that $\mathcal B^{\vphantom{\dagger}}(\mathcal B^{\dagger})$ acting on the normalized $|N_B\rangle$ gives us $\sqrt{N_B(NM-2N_B)} |N_B-1\rangle \bigl(\sqrt{(N_B+1)(NM-2N_B)}|N_B+1\rangle \bigr)$. Then, we can write down the master equation accordingly, 
and the solution would have the same structure as in $M=1(U/t\neq 0)$.

We compare the analytical solution with the numerics in Fig.~\ref{fig:U=0}. The alignment between them confirms our arguments above.

\section{Derivation of $M=1$ density matrices and observables}
We first consider the Mott phase, where $r_<=I(\mu-t)/\gamma(\mu-t) \ll 1$ or $r_>=I(\mu+t)/\gamma(\mu+t)\gg 1$. For $r_< <1$, from Eq.~(26) in the main text, we have

\begin{equation}
\begin{aligned}
    \rho_m
    &=\rho_0\prod_{j=1}^m r(\omega_j) 
    \notag \\
    &=\rho_0 r_<^m \prod_{j=1}^m \bigl(1+\frac{r_<'}{r_<}\frac{2tj}{N}\bigr) 
    \notag \\
    &\approx \rho_0 r_<^m \exp{\biggl(\sum_{j=1}^m \frac{2r_<'t}{r_<N}j\biggr)} 
    \notag \\ 
    &\approx \rho_0 r_<^m \exp{\biggl( \frac{r_<'t}{r_<N}m^2\biggr)}
\end{aligned}
\end{equation}
where we Taylor expand $r_{j}$, convert the product to the exponential of a sum, and keep only the leading order term in a $1/N$ expansion of the exponential. The weight $\rho_m/\rho_0$ at large $m$ is suppressed by $r_<^m$ as well as the exponential function. This calculation is controlled because $m\lesssim \sqrt{N}$ are the only values where $\rho_m$ is substantial (if $r_<\le 1$).  

When $r_< \ll 1$, $\rho_m$ decays very fast and we are in the deep Mott phase. Note that $r_<'<0$ is exactly our  assumption that $r(\omega)$ monotonically decreases, as stated in the text. Similarly, we can consider the case $r_> \gg 1$, and we get
\begin{equation}
    \rho_{N-m}=\rho_N \prod_{j=0}^{m-1} [r(\omega_{N-j})]^{-1} \approx \rho_N r_>^{-m} \exp{\biggl(\frac{r_>'t}{N}m^2\biggr)}
\end{equation}

Next, we analyze the intermediate phase, when $r_< >1 > r_>$ and $r_{j^*}=1$ with $j^*/N\sim O(1)$. 
An almost identical calculation gives
\begin{equation}
\begin{aligned}
    \rho_{j^*+m} &= \begin{cases} \rho_{j^*} \prod_{j=0}^m \left(1+r_{j^*}' \frac{2tj}{N}\right) & m\ge 0 \\
    \rho_{j^*} \prod_{j=0}^{-m} \left(1-r_{j^*}' \frac{2tj}{N}\right)^{-1}& m\le 0
    \end{cases} 
    \notag \\
    &\approx
     \rho_{j^*} \exp{\biggl( \frac{r_{j^*}'t}{N}m^2\biggr)}
\end{aligned}
\end{equation}
where we used $r_{j^*}=1$, and thus we have only an exponential decay term compared to the Mott phase, which would be slower since the exponent is suppressed by $1/N$. This holds true until $m^2$ reaches $N$, where the density matrix weight is already negligible. Therefore, we should have $m\lesssim \sqrt{N}$. For $j^*$ near $0$ or $N$, $j^*\pm m$ should be truncated by $0$ and $N$, respectively.

Now, we can calculate the observables and critical behaviors. Deep inside the superfluid phase, we have the order parameter $\Psi^2=\langle \mathcal B^{\dagger} \mathcal B^{\vphantom{\dagger}}\rangle /N^2$
\begin{align}
    \Psi^2
    &=\frac{\sum_m (j^*+m)(N-j^*-m+1) \exp{\bigl(\frac{r_{j^*}'t}{N}m^2\bigr)}}    {N^2\Bigl(\sum_m \exp{\bigl(\frac{r_{j^*}'t}{N}m^2\bigr)}\Bigr)}  
    \notag \\
    &\approx \frac{j^*}{N}\left(1-\frac{j^*}{N}\right) + O(1/\sqrt{N})
\end{align}
where  $m$ is summed from $-j^*$ to $N-j^*$ but only the terms with $|m| \lesssim \sqrt{N}$ contribute. We keep only the leading order terms.

From the Mott phase side, we can calculate $\Psi^2$ as well. We assume $r_< \ll 1$, we can ignore the exponential 
corrections and evaluate 
\begin{equation}
\begin{aligned}
    \Psi^2 &=
    \frac{\sum_{m=0}^N m(N-m+1)r_<^m \exp{\bigl(\frac{r_<'t}{r_< N}m^2\bigr)}}{N^2 \Bigl(\sum_{m=0}^N r_<^m \exp{\bigl(\frac{r_<'t}{r_< N}m^2\bigr)}\Bigr)} 
    \notag \\
    &\approx
    \frac{1}{N} \Bigl(\frac{\sum_m m r_<^m}{\sum_m r_<^m}\Bigr) \notag \\
    &=\frac{1}{N} \frac{r_<}{1-r_<} 
\end{aligned}
\end{equation}
A similar calculation for $r_> \gg 1$ holds and we find $\Psi^2 \approx (1-r_>^{-1})^{-1}/N$.

Finally, if we approach the phase transition from the left, $r_<\to 1^{-}$, we use $r_<^m =e^{\ln(r_<^m)}\approx e^{m(r_<-1)}$ to write
\begin{equation}
\begin{aligned}
    \Psi^2 
    &\approx \frac{1}{N^2} 
    \frac{\rho_0\sum_{m=0}^{N}m(N-m+1) e^{\frac{r_<'t}{r_<}\frac{m^2}{N} + m(r_<-1)}}{\rho_0 \sum_{m=0}^{N}e^{\frac{r_<'t}{r_<}\frac{m^2}{N} + m(r_<-1)}}\notag \\
    &\approx  \frac{1}{N^{1/2}}
    \frac{\int_0^{\sqrt{N}}dx x \exp{\left[\frac{r_<'t}{r_<}x^2-(1-r_<)\sqrt{N}x\right]}}{\int_0^{\sqrt{N}}dx \exp{\left[\frac{r_<'t}{r_<}x^2-(1-r_<)\sqrt{N}x\right]}}.
\end{aligned}
\end{equation}
where we changed to the variable $x=m/\sqrt{N}$ and approximated the sums by integrals. The integrand quickly decays when $x\gtrsim 1$ so we can extend the limits of integration to infinity, and 
we then get the scaling form $\Psi^2 =N^{-1/2}f_s(N^{1/2} (1-r_<))$. We verified that the same scaling form applies when we approach the phase transition from either side, and $n_B=\langle N_B \rangle/N$ has a scaling form with the same critical exponents as well.

\section{Extrapolation details}
\begin{figure}
\centering    \includegraphics[width=0.45\textwidth]{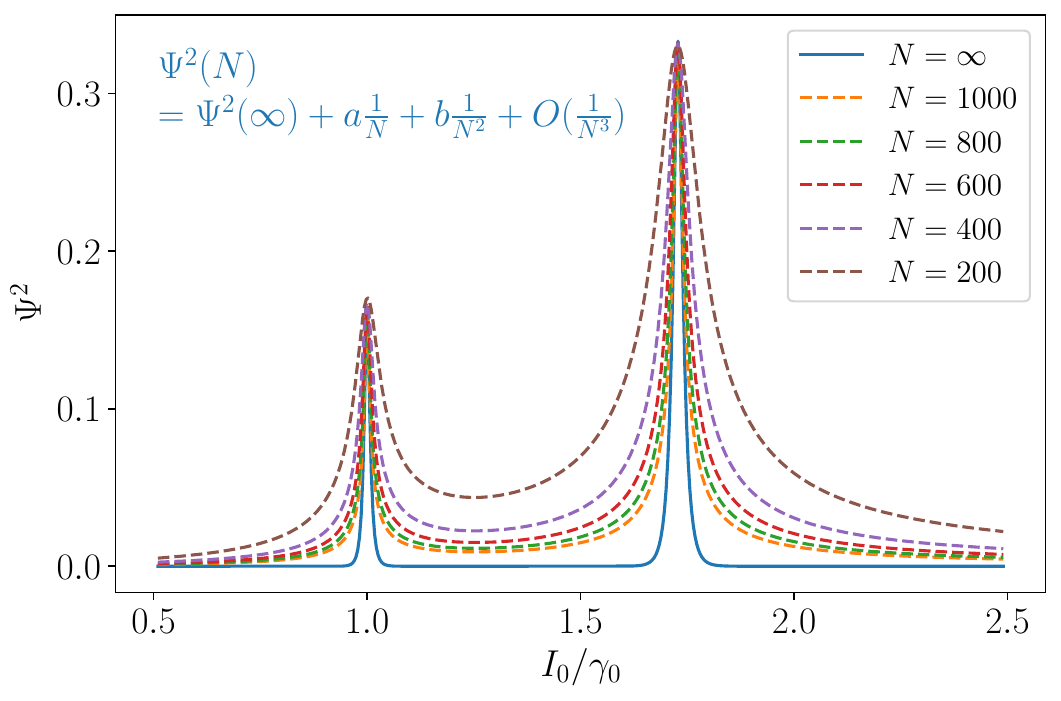}
    \caption{The details of extrapolation to $N=\infty$ for $\Psi^2$ for $M=2$ at $t=0, U=30, \mu=150$. The same procedures are carried out for other $t/U$ values and other observables.}
    \label{fig:extrap detail}
\end{figure}
In our numerical calculations, the lattice size can be up to 1000 for $M=2$ and 100 for $M=3$, within our computational capability. To get the phase diagram shown in the text, we extrapolate the data at each point of $t/U$ and $I_0/\gamma_0$ of several system sizes to thermaldynamic limit ($N=\infty$). 
We use a polynomial extrapolation of order 2 (quadratic fitting), which is equivalent to assuming $O(N)=O(N=\infty)+O' \frac{1}{N} + O'' \frac{1}{N^2} + ...$ and truncate it up to the second order of $1/N$. We find that the truncation up to quadratic order already gives us fair results. The fitting detail is shown in Fig.~\ref{fig:extrap detail}, where not all finite $N$ curves are plotted. For the extrapolation of the entropy, we get negative extrapolated entropy values at negligibly few points. We manually set them to be 0 since a negative entropy is not physical and we believe this abnormal behavior is a numerical artifact during the extrapolation.
We also tried a power law fitting in $1/N$, which gives almost the same results but fails near the phase transitions. We adopt the polynomial fitting results for it gives a more smooth behavior globally.

\section{Phase diagrams}
\begin{figure*}
\centering    
\includegraphics[width=0.8\textwidth]{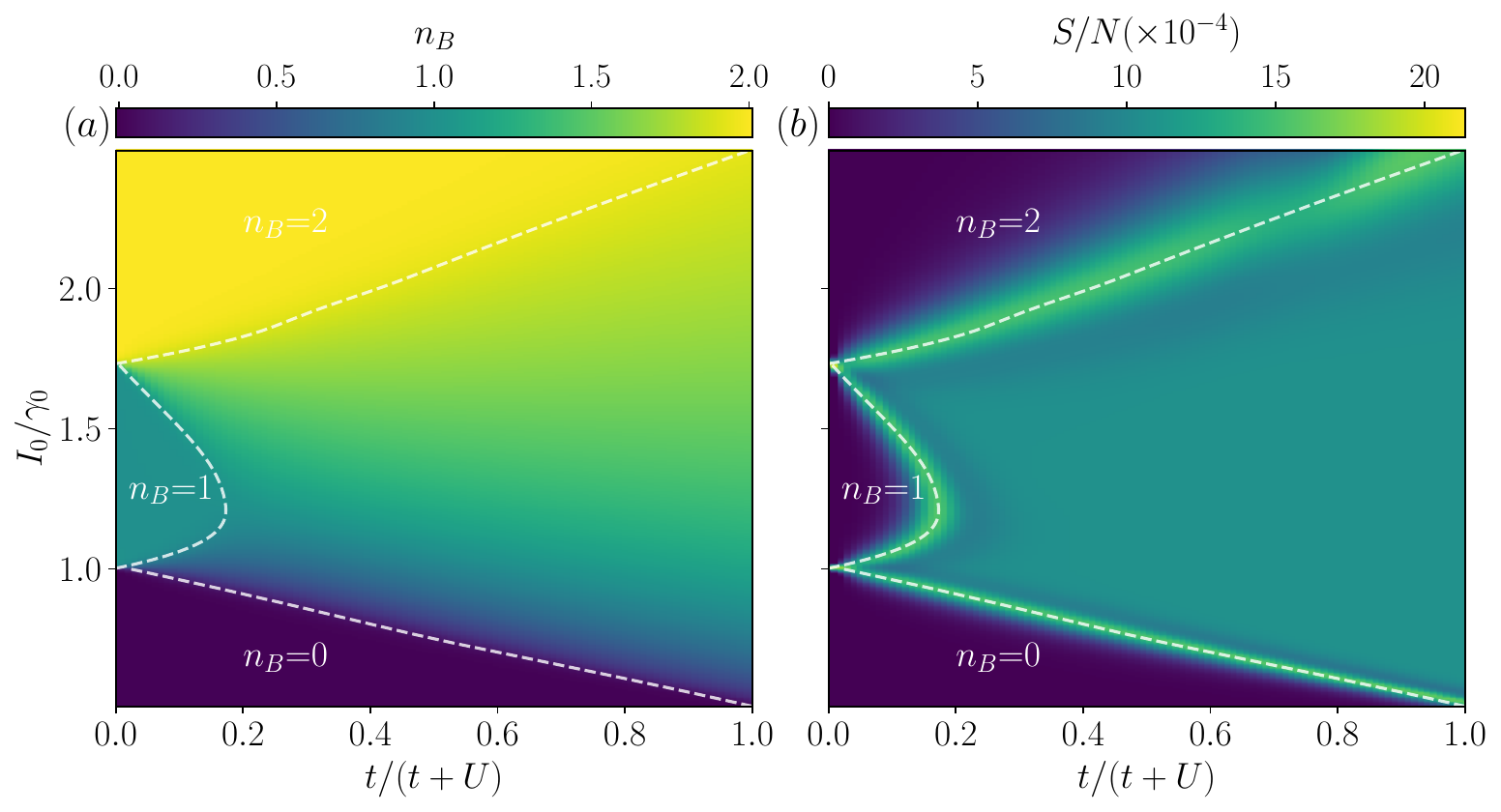}
    \caption{The phase diagrams mapped in terms of (a) $n_B$ and (b) $S/N$, for $M=2$. Every data point is extrapolated to $N=\infty$ as indicated in the Supporting text.}
    \label{fig:M2 phase diagram}
\end{figure*}

\begin{figure*}
\centering    
\includegraphics[width=0.8\textwidth]{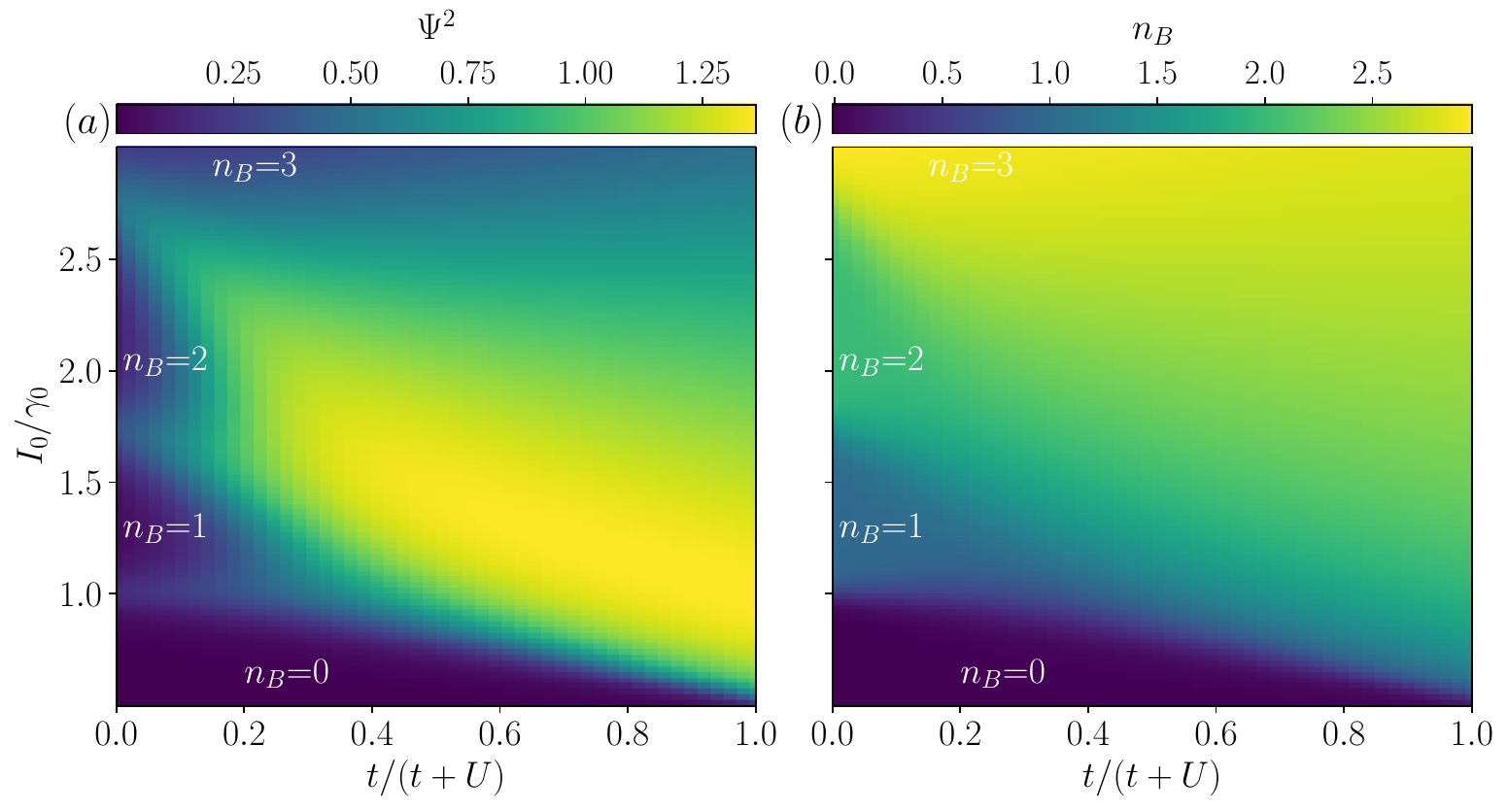}
    \caption{The phase diagrams mapped in terms of (a) $\Psi^2$ and (b) $n_B$, for $M=3$. We see the Mott lobes appearing; although the $n_B=2$ region has now become a lobe, the $n_B=1$ lobe seems almost identical.}
    \label{fig:M3 phase diagram}
\end{figure*}
In the text we showed the phase diagram in terms of $\Psi^2$. Here we show the phase diagrams mapped in terms of $n_B$ and $S/N$ in Fig.~\ref{fig:M2 phase diagram}. We see that the extracted phase boundaries from the entropy clearly delineate the Mott regions seen in the $n_B$ plot.

We also show the phase diagram for $M=3$, by plotting $\Psi^2$ and $n_B$ vs. $I_0/\gamma_0$ and $t/(t+U)$ in Fig.~\ref{fig:M3 phase diagram}. In this case, we can only reach up to $N=70$, but we can already see the Mott insulating domes appearing. The shape of the $n_B=2$ lobe is clearly very different from the $M=2$ case, but the $n_B=1$ lobe appears very similar. This plot demonstrates that the effect of the truncation is unlikely to affect the physics near the $n_B=1$ lobe.

\section{Critical scaling }
\subsection{Mott-Mott transition}
\begin{figure}
    \centering    \includegraphics[width=0.45\textwidth]{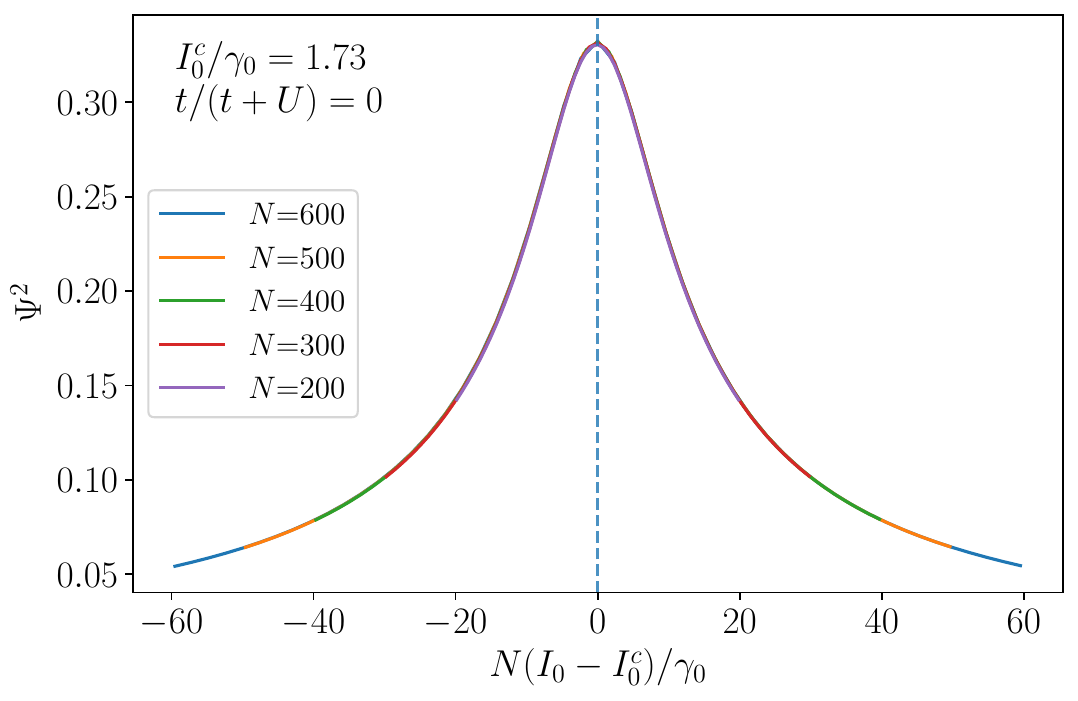}
    \caption{Scaling collapses for $\Psi^2$ at the Mott-Mott transition.}
    \label{fig:corr_collapse_I}
\end{figure}
In the main text, we studied $n_B$, the order parameter for characterizing the Mott-Mott transition point, and confirmed the critical scaling form $n_B=f_{n_B,m}(N(I_0-I_c))$. In Fig.~\ref{fig:corr_collapse_I} we show the scaling collapse for $\Psi^2$ at this transition point, and the same scaling form as $n_B$ also applies here i.e.~$\Psi^2=f_{\Psi^2,m}(N(I_0-I_c))$.

\subsection{Mott-Superfluid transition}
\begin{figure}
    \centering    \includegraphics[width=0.45\textwidth]{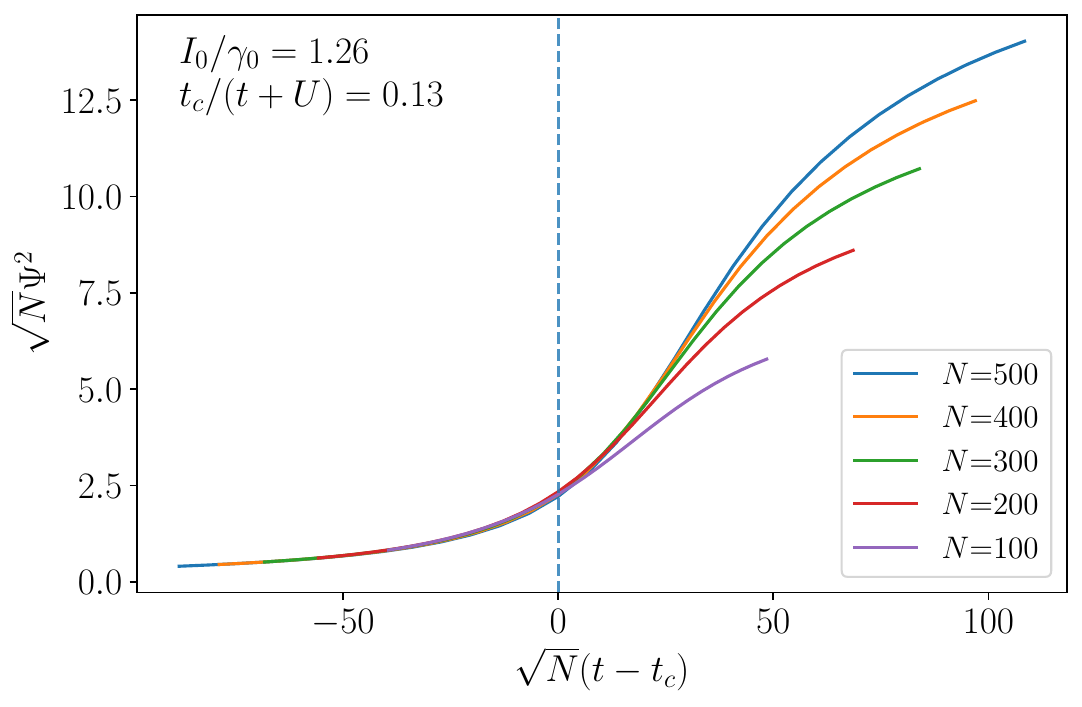}
    \caption{Scaling collapses at the Mott-superfluid transition, where $t/U$ is tuned and $I_0/\gamma_0$ is fixed.}
    \label{fig:collapse_t}
\end{figure}
In the main text, we displayed the critical scaling from Mott to superfluid phases by tuning $r=I_0/\gamma_0$, where we confirmed the scaling form $\Psi^2(N)=N^{-1/2}f_{\Psi^2,s}(\sqrt{N}(r-r_c))$. In Fig.~\ref{fig:collapse_t} we show that this scaling form applies if we tune $t/U$ at fixed $I_0/\gamma_0$ instead, and, in both cases, the same exponents are found. 

We expect that this scaling form should almost everywhere at the Mott-superfluid phase boundary. In the equilibrium case, there are two different universality classes for the Mott-superfluid transition depending on whether the density remains at $n_B\in \mathbb Z$ (at the peak of the Mott lobe) or if the density is generic \cite{SachdevQPT}. The two can be distinguished as having different values of the dynamic critical exponent, $z$. In principle, the Liouvillian gap, described in the next section, determines how quickly the system approaches the steady state and could probe $z$. In practice, we can only reach the scaling regime for the Liouvillian gap for $M=1$, which does not have the ``special'' critical point as there is no peak of the Mott lobe.

\subsection{Liouvillian gap}
\begin{figure*}
    \centering    \includegraphics[width=\textwidth]{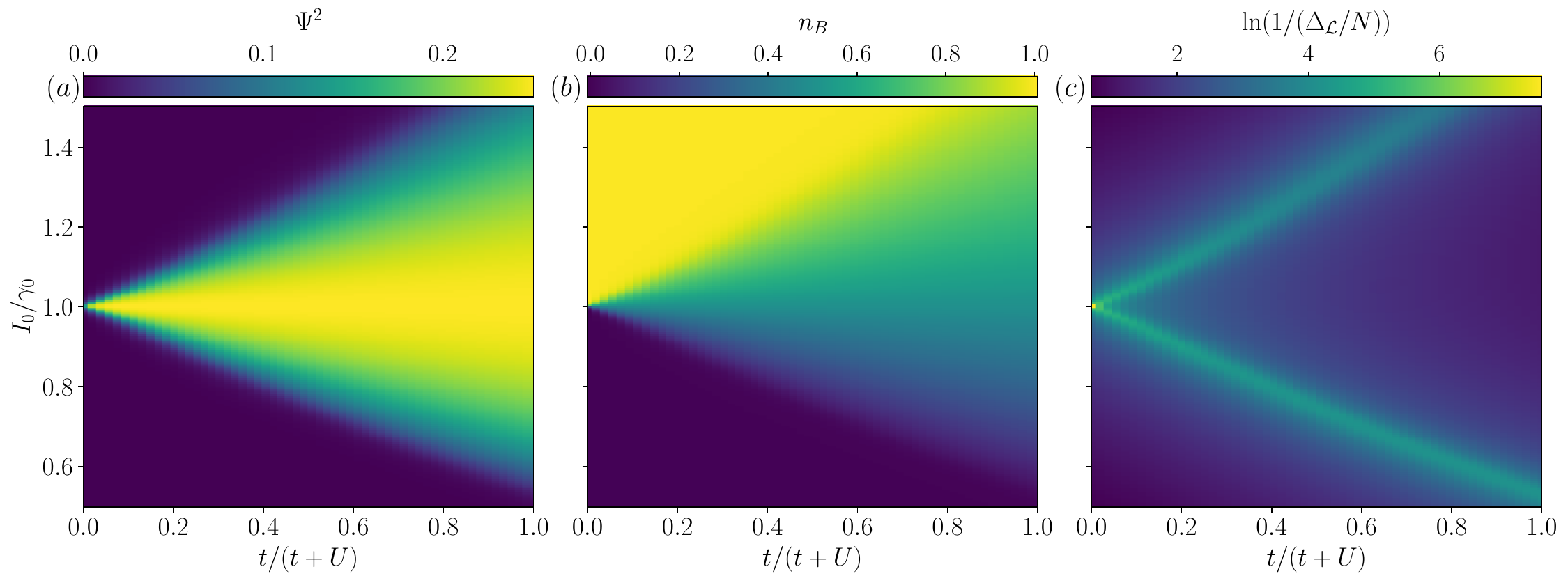}
    \caption{We plot (a) $\Psi^2$, (b) $n_B$ and (c) the inverse Liouvillian gap on a log scale, for $M=1$. We see that the Liouvillian gap is smallest at phase boundaries.}
    \label{fig:Liouv gap}
\end{figure*}
The Liouvillian, $\mathcal L$, is the superoperator defining the right-hand-side of the master equation $\dot \rho = \mathcal L \rho$. Here we measure the Liouvillian gap, $\Delta_{\mathcal L}$ defined as the smallest real part of the non-zero eigenvalues of $\mathcal L$ \cite{Zhang2022}. This gap determines the characteristic time for the system to reach its steady state. In Fig.~\ref{fig:Liouv gap}, it's clear that $\Delta_{\mathcal{L}}$ reaches a minimum at the (Mott-superfluid) phase boundary, which implies a critical slowing down near the phase transitions. 
 As in the superradiance phenomenon, the system decay rate scales with $N$\cite{breuer2002theory}, so we rescaled $\Delta_{\mathcal L}/N$ to make it clear that there is a slowing down at the phase transition relative to the rest of the phase diagram. 
 
 \begin{figure*}
    \centering    \includegraphics[width=0.45\textwidth]{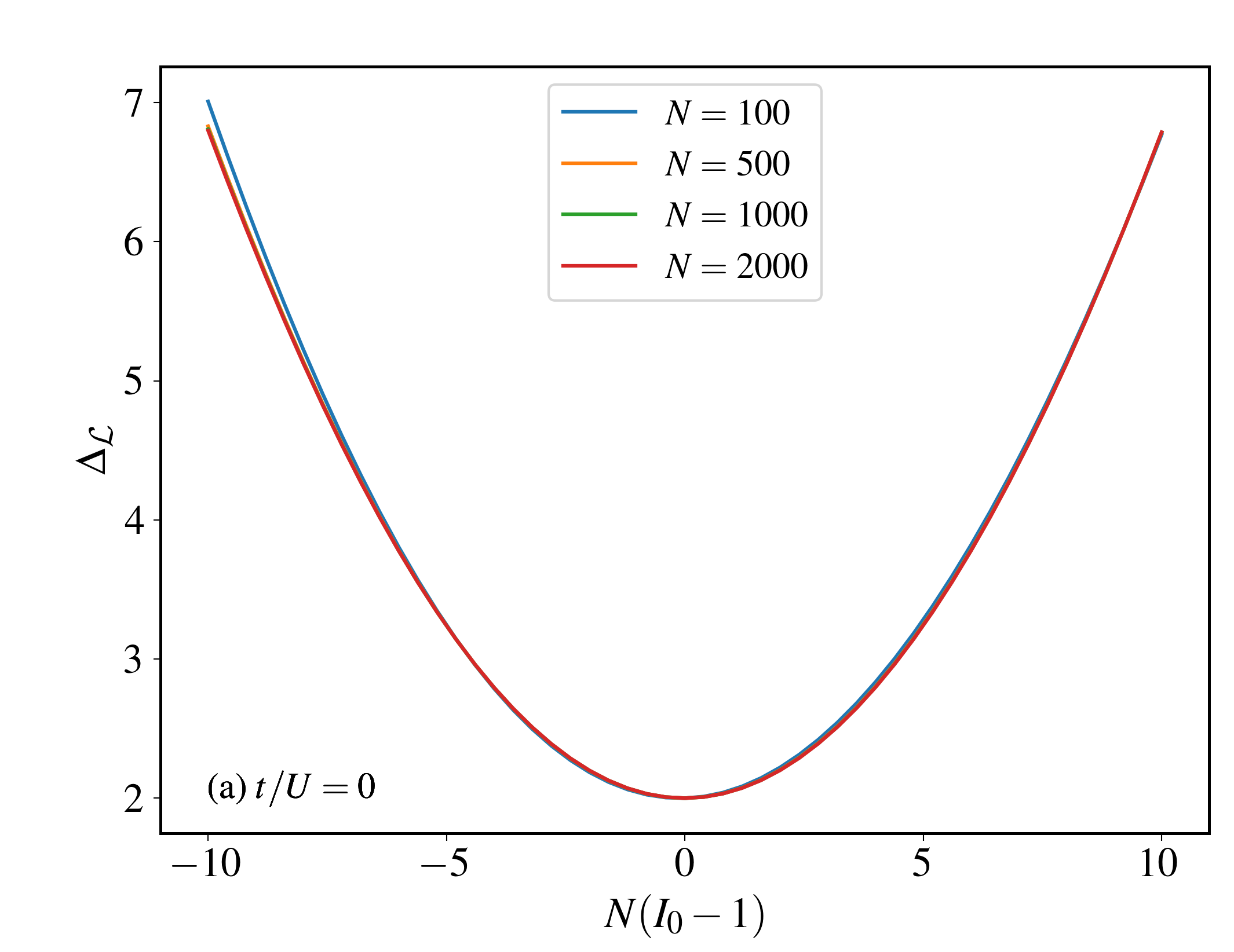}
    \includegraphics[width=0.45\textwidth]{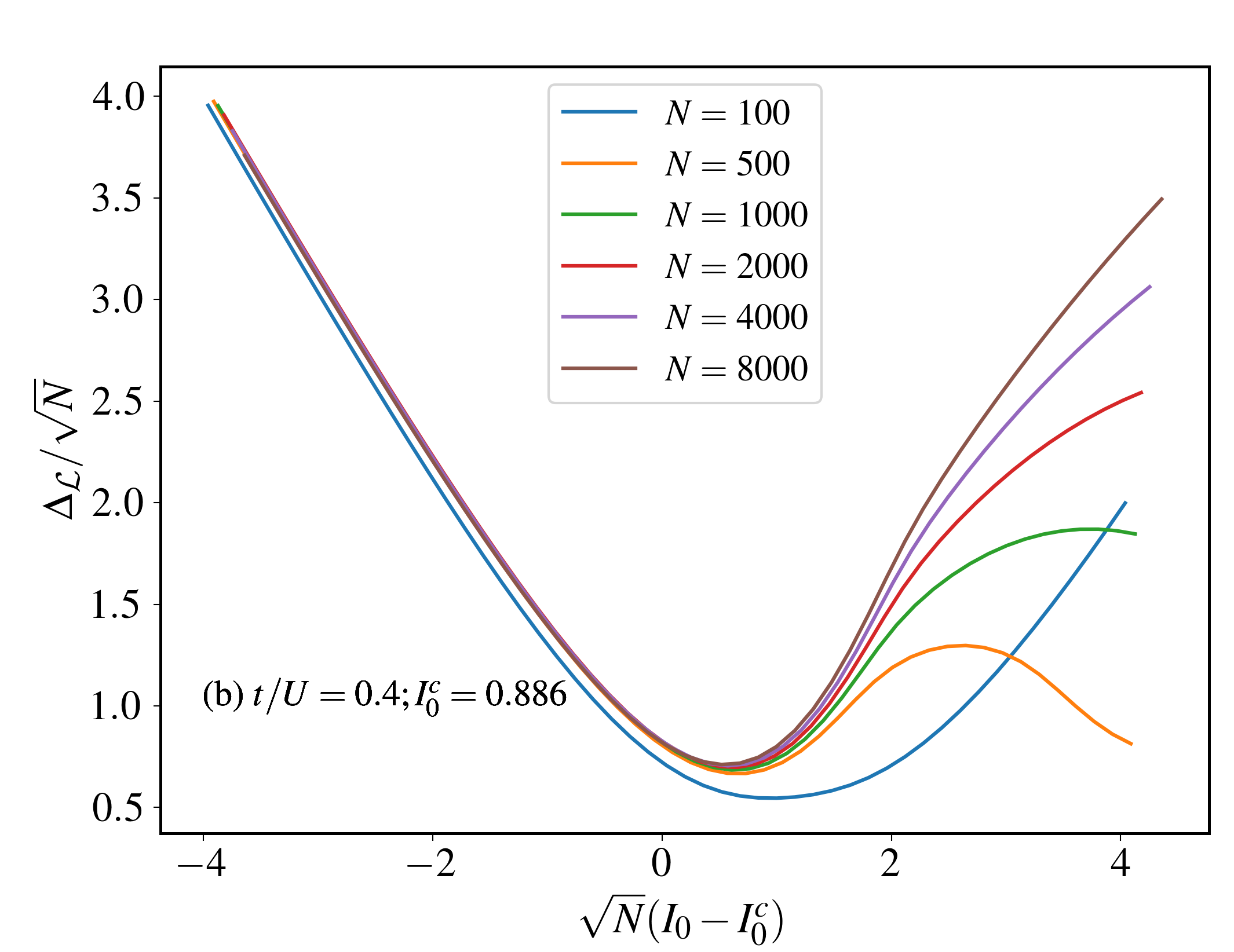}
    \caption{We plot the Liouvillian gap scaling for the Mott-Mott (a) and Mott-superfluid (b) phase transitions for the $M=1$ case. In both cases $\Delta_{\mathcal L}\sim N$ away from the critical point as in the superradiance case. The scaling collapse for the Mott-superfluid case has more finite-size effects due to the superfluid phase being an intermediate phase between two Mott phases. }
    \label{fig:Liouville_collapse}
\end{figure*}
 In Fig.~\ref{fig:Liouville_collapse}, we perform a scaling collapse for $\Delta_{\mathcal L}$ with $M=1$. In equilibrium, the energy gap $\Delta E \sim (t-t_c)^{\delta}$ where $\delta=\nu z$ for $z$ the dynamic critical exponent and $\nu$ the correlation length exponent. In the mean-field theory of the equilibrium Bose-Hubbard model $\delta=1$. Similarly, we see that $\Delta_{\mathcal L}/N = N^{-\delta'/\lambda} g_{\mathcal L}(N^{1/\lambda}(I_0-I_c))$ for some scaling function $g_{\mathcal L}$ and exponent $\delta',\lambda$ (that are potentially different between the Mott-Mott and Mott-superfluid transition) implying that $\Delta_{\mathcal L}/N \sim (I_0 -I_c)^{\delta'}$ close to the transition.  We numerically observe $\delta_s'=\delta_m'=1$ (as well as $\lambda_m=\lambda_s-1=1$ as before) similar to $\delta_s=1$ for the equilibrium transition. We consider $\Delta_{\mathcal L}/N$ instead of $\Delta_{\mathcal L}$ so that a finite value is reached away from the critical point in the thermodynamic limit.
 
Calculations of $\Delta_{\mathcal L}$ for $M\ge 2$ are beyond our computational capabilities, but we expect that the slowing-down and scaling form do not depend on the truncation $M$ and will hold for general $M$.

\section{Steady state distributions}
\begin{figure*}
    \centering    \includegraphics[width=0.45\textwidth]{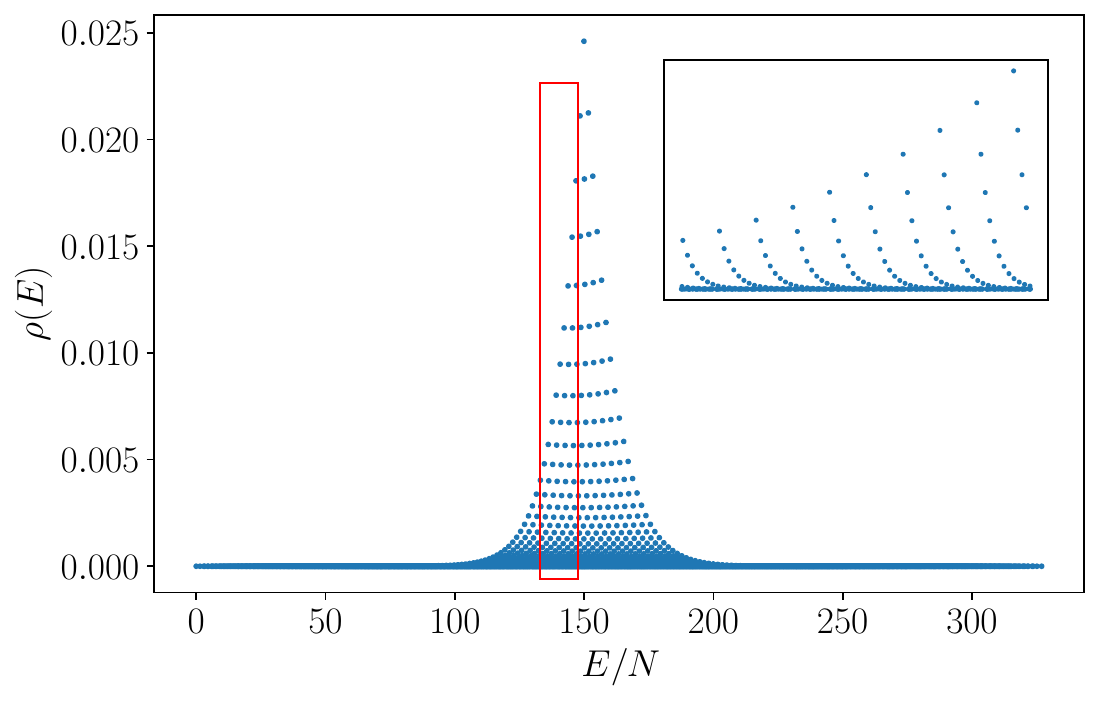}
    \caption{The non-equilibrium steady state density matrix distribution for $M=2$, at $I_0/\gamma_0=1.26, t/(t+U)=0.1$. The inset is the zoomed-in region in the red rectangle.}
    \label{fig:non_equil_distrib}
\end{figure*}
It's worth noting that in our model, the resulting steady states are non-thermal. In Fig.~\ref{fig:non_equil_distrib}, we display the steady state density matrix distribution with respect to system energy obtained by numerics at a generic point in the phase space ($I_0/\gamma_0=1.26, t/(t+U)=0.1$), and $M=2$. In the inset, we can see a oscillation feature within each $N_B$ sector, which emphasizes the non-thermal property of the steady state. The peak structure of $\rho(E)$ cannot be explained through a grand canonical form for $\rho(E)\sim e^{-\beta(E-\mu N)}$.

\section{Comparison with equilibrium system}
\begin{figure*}
    \centering    \includegraphics[width=0.8\textwidth]{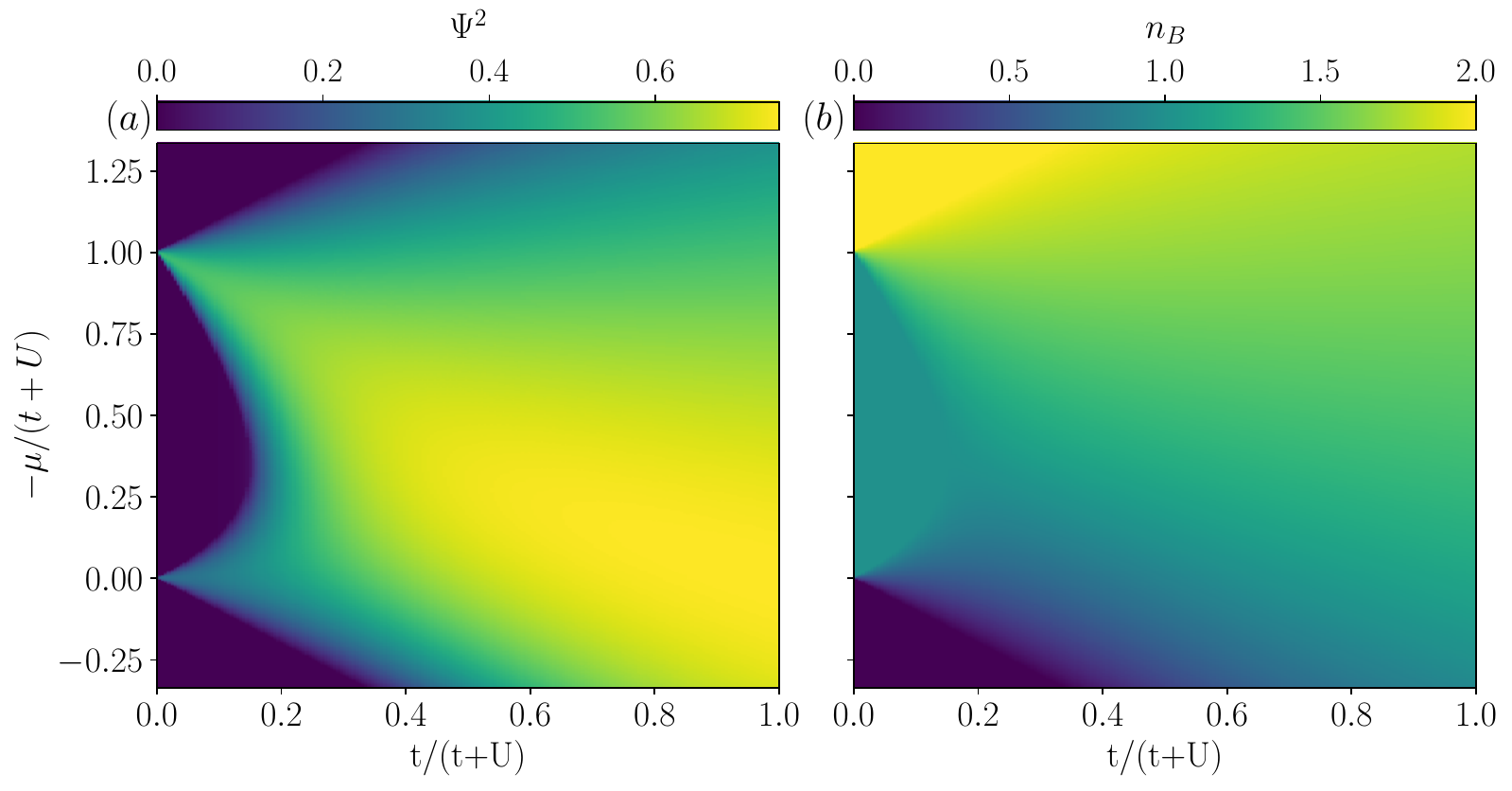}
    \caption{The (zero-temperature) equilibrium phase diagrams for the all-to-all hopping Bose Hubbard model. The scaled observables $\Psi^2$ and $n_B$ do not depend on $N$.}
    \label{fig:equil phase}
\end{figure*}
In Fig.~\ref{fig:equil phase}, we show the (zero-temperature) equilirbium phase diagram of the all-to-all hopping Bose Hubbard Hamiltonian, in terms of $t/(t+U)$ and $\mu/(t+U)$, where $\mu$ is the bosonic chemical potential. The nonequilibrium and equilibrium phase diagrams have similar shapes, which agrees with our intuition that increasing the intensity of the light is analogous to changing the chemical potential.

However, there are key differences: First, at $t/(t+U)=0$, the order parameter $\Psi^2$ remains zero at any $\mu$. This is different than the non-equilibrium case, where $\Psi^2$ has a peak and shows critical scaling at the transition point of two Mott lobes. 
Second, at any point in the phase diagram, the observables ($\Psi^2=\langle \mathcal{B}^{\dagger} \mathcal{B}^{\vphantom{\dagger}}\rangle/N^2$ and $n_B=\langle N_B \rangle/N$) do not scale with $N$ when $N$ is large enough, which is reasonable since the all-to-all hopping is equivalent to enforcing a mean field behavior. 
Third, the entropy remains zero in the whole phase diagram, because at zero temperature, we are always at the ground state, which in general has no degeneracy. However, we observe a peak of entropy at the non-equilibrium phase transition.

\begin{figure*}
    \centering \includegraphics[width=0.85\textwidth]{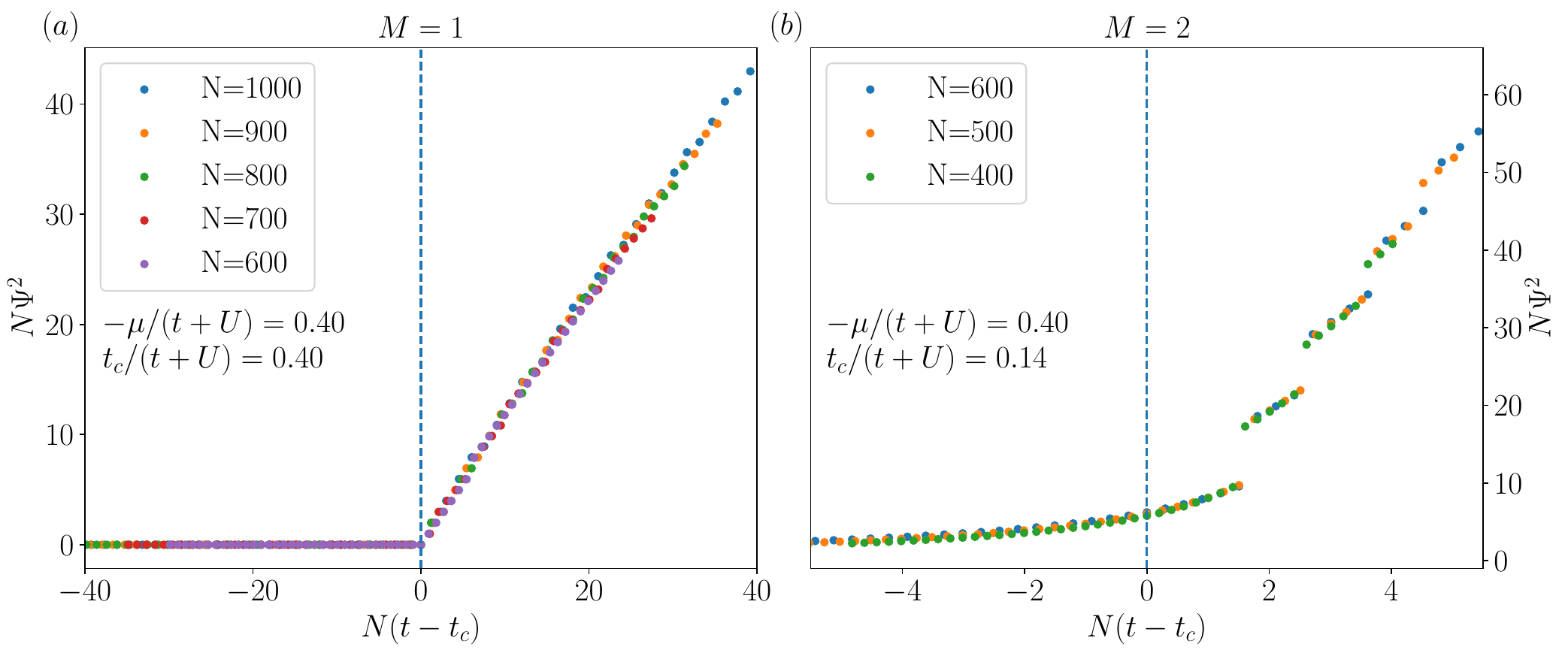}
    \caption{We plot the scaling collapse for $\Psi^2$ for (a) $M=1$ and (b) $M=2$ respectively, for the  transition from the Mott phase ($n_B=1$) to the superfluid phase. We see 
    that the equilibrium transition has $\lambda_s = 1$ as opposed to the nonequilibrium transition with $\lambda_s=2$.}
    \label{fig:SM_eqscalingcollapse}
\end{figure*}

Finally, we comment on the critical exponents. For the equilibrium Mott-Mott transition, we can only tune $\mu$ to move directly from e.g. the $n_B=0$ to the $n_B=1$ lobe. The transition is simply between the eigenstate with every site having $0$ bosons and the eigenstate with every site having $1$ bosons. By analyzing the energy per site, it is clear that this transition is first-order. In the nonequilibrium case, we find that this transition has become second order with an order parameter, e.g. $n_B = \Theta(I_0 - I_c)$ for the $n_B=0$ to $n_B=1$ lobe, that has the finite-size scaling form $n_B = N^{-\beta_m/\lambda_m} f_{n_B,m}(N^{1/\lambda_m}(I_0-I_c))$ with $\lambda_m=\beta_m+1=1$.

For the Mott-superfluid transition, as we mentioned above, there is a ``generic'' transition, occurring in most of the phase diagram, and a ``special'' transition occurring at the peak of the Mott lobe. Both critical points have an energy gap that scales as $\Delta E \sim (t-t_c)^{\nu z}$ with $\nu z = 1$ but the former has $z=2$ and the latter $z=1$. We are unsure if we can capture the ``special'' transition in our nonequilibrium model, but we can determine the critical exponents $\beta_s$ and $\lambda_s$ as before for the generic case. 

Our all-to-all model mimics the mean-field limit of the Bose-Hubbard model \cite{Fisher1989}, and we can then immediately know that the mean-field value of $\beta_s = 1$ (since our order parameter is $\Psi^2$ and not $\Psi$). 
In fact, we can predict the scaling behaviors through a path integral formulation. The partition function, with an applied external field ($-h^*b_i^{\vphantom{\dagger}}-h^{\vphantom{*}}b_i^{\dagger}$), is given by
\begin{widetext}
\begin{align}
    Z(h)
    &=\int \mathcal Db_i^{\vphantom{*}}\mathcal Db_i^*
    e^{-\int d\tau\biggl(-\frac{t}{N}\bigl(\sum_i b_i^*\bigr)\bigl(\sum_j b_j^{\vphantom{*}}\bigr)
    +\sum_i \bigl(b_i^* \partial_{\tau}b_i^{\vphantom{*}}+Un_i^2-\mu n_i\bigr)
    -\sum_i\bigl(h^*b_i^{\vphantom{*}}+h^{\vphantom{*}}b_i^*\bigr)
    \biggr)}
    \\
    &=\int \mathcal Db_i^{\vphantom{*}}\mathcal Db_i^* \mathcal D \psi^{\vphantom{*}} \mathcal D \psi^*
    e^{-\int d\tau\biggl(
    \sum_i\Bigl((\psi^*-h^*)b_i^{\vphantom{*}}+(\psi-h)^{\vphantom{*}}b_i^* +b_i^* \partial_{\tau}b_i^{\vphantom{*}}+ Un_i^2-\mu n_i\Bigr)
    +\frac{N}{t}|\psi|^2
    \biggr)}    
\end{align}
\end{widetext}
In the second line, we make a Hubbard-Stratonovich transformation to decouple the hopping term, and only a single auxiliary field is needed due to the all-to-all hopping form. Since all sites are decoupled, we can integrate out each $b_i$ and arrive at an effective action for field $\psi$, where each $b_i$ integration gives the same contribution.

\begin{widetext}
\begin{align}
    S_{\text{eff}}
    &=
    \int d\tau
    \Bigl(
    Nf(\psi-h)-\frac{N}{t}|\psi^2|
    \Bigr)
    \\
    &=    \int d\tau
    N\Bigl(
    r_0|\psi-h|^2+u_0|\psi-h|^4-\frac{1}{t}|\psi^2|+...
    \Bigr)
    \\
    &=\int d\tau
    N\Bigl(
    r_0(|\psi|^2-h^*\psi-h\psi^*)+
    u_0(|\psi|^4-2h^*\psi|\psi|^2+\text{c.c.})
    -\frac{1}{t}|\psi^2|+...
    \Bigr)
    \\
    &=\int d\tau
    N\Bigl(
    r|\psi|^2+u_0|\psi|^4
    -r_0h\psi^*-2u_0h\psi^*|\psi|^2+\text{c.c.}+...
    \Bigr)
\end{align}
\end{widetext}
where $f(x)$ is some function obtained by integrating out $b_i$ field. In the second line we Taylor expand $f(x)$ according to symmetry constraints, and we ignored the Berry phase term $\psi^*\partial_{\tau}\psi^{\vphantom{*}}$ as we assume that time-independent fields are most important at the saddle point \cite{Fisher1989}. In the third line we expand $|\psi-h|$ and omit terms of higher orders in $h$ or eventually $1/N$. We define $r\equiv r_0-1/t$, and at the critical $t_c$, $r$ should go to zero, and $r\sim \delta t=t-t_c$ to the lowest order.
Now, we rescale $\psi \to \psi/\sqrt{Nr}$. If there is any change in the measure of $\int \mathcal D \psi$, it only contributes a constant term $C(N)$ to the free energy.
Then, we have
\begin{widetext}
\begin{align}
    S_{\text{eff}}
    &=\int d\tau
    \Bigl(
    |\psi|^2+\frac{u_0}{Nr^2}|\psi|^4
    -\frac{r_0\sqrt{N}}{\sqrt{r}}h\psi^*
    -\frac{2u_0}{\sqrt{Nr^3}}h\psi^*|\psi|^2+\text{c.c.}+...
    \Bigr)
    \\
    &\sim
    \int d\tau
    \Bigl(
    |\psi|^2+\frac{u_0}{N\delta t^2}|\psi|^4
    -\frac{r_0\sqrt{N}}{\sqrt{\delta t}}h\psi^*
    -\frac{2u_0}{\sqrt{N\delta t^3}}h\psi^*|\psi|^2+\text{c.c.}+...
    \Bigr)
\end{align}
\end{widetext}
So, the free energy $f=-\frac{1}{\beta N}\ln Z=-\frac{1}{\beta N}\int \mathcal D\psi\mathcal D \psi^* e^{-S_{\text{eff}}}$ should take the form of
\begin{align}
    Nf(N,h)=F\bigl(N\delta t^2,h\sqrt{\frac{N}{\delta t}},\frac{h}{\sqrt{N\delta t^3}}\bigr)
\end{align}
And we can calculate the order parameter $\Psi^2$ at zero external field as
\begin{align}
    \Psi^2&=
    \langle \bigl(\sum_ib_i^{\dagger}\bigr)\bigl(\sum_jb_j\bigr)\rangle/N^2
    \\
    &=
    \frac{1}{N^2}\frac{\partial^2}{\partial h \partial h^*}
    F\bigl(N\delta t^2,h\sqrt{\frac{N}{\delta t}},\frac{h}{\sqrt{N\delta t^3}}\bigr)\Bigl|_{h=0}
    \\
    &=\frac{G_1\bigl(N\delta t^2\bigr)}{N\delta t}
    +\frac{G_2\bigl(N\delta t^2\bigr)}{(N\delta t)^2}
    +\frac{G_3\bigl(N\delta t^2\bigr)}{(N\delta t)^3} 
\end{align}
where $G_1(x)$, $G_2(x)$ and $G_3(x)$ are different unknown functions. When we perform finite-size scaling collapses, we fix the lattice size $N$, and examine $\Psi^2(N)$ near the critical point, i.e.~$\delta t\sim0$. Therefore, we should look at $\Psi^2$ at the limit $\delta t\to 0$. 

Because $\Psi^2\to 0$ at the transition, we must have $G_1(0)=G_2(0)=G_3(0)=0$. If we assume that $G_i(x)$ can be Taylor expanded at $N\delta t^2=0$, we find $G_3'(0)=0$ and $\Psi^2 = G_1'(0) \delta t + \frac{1}{N}G_2'(0) + ...$. Equivalently, $N\Psi^2\sim G_1'(0) N\delta t+G_2'(0)+...$, which is consistent with the scaling collapse we find numerically in Fig.~\ref{fig:SM_eqscalingcollapse} for $M=1$ and $M=2$.  We thus confirm that $\lambda_s=\beta_s=1$ at the ``generic'' transition point in the equilibrium case, in contrast to the nonequilibrium $\lambda_s=\beta_s+1=2$.

As a double check, we can calculate the scaling forms in the case of $M=1$ where an exact solution can be obtained. Recall that there is only one state in the (fully symmetrized) sector of $N_B$ bosons, labeled as $|N_B\rangle$. We have
\begin{equation}
\begin{aligned}
    E_{N_B} &= \mu N_B - \frac{t}{N} N_B (N-N_B+1);
    \notag \\
    \frac{\mathcal B^\dagger \mathcal B}{N^2} |N_B\rangle &= \frac{N_B}{N}(1-\frac{N_B}{N}+\frac{1}{N}).
\end{aligned}
\end{equation}
By treating $n_B=N_B/N$ as a continuous variable, it is easy to find that the Mott-superfluid transition occurs at $t_c=-\mu/(1-\frac{1}{N})\sim -\mu$, where we choose $\mu<0$ to study the $n_B=1$ to superfluid phase transition. For the $\mu>0$ transition, the same analysis can be carried out. The order parameter can be obtained as
\begin{align}
    \Psi^2=
    \begin{cases}
        \frac{1}{N} &  t<t_c
        \\
        \frac{1}{4}-\frac{\mu^2}{4t^2}+\frac{1}{2N}+\frac{1}{4N^2} &  t>t_c 
    \end{cases}
\end{align}
And we can expand $t=\delta t-\mu$ near the critical point. After keeping only the lowest order terms of $\delta t$ (suitable for doing finite size scaling), we get
\begin{equation}
    \Psi^2
    \sim -\frac{\delta t}{2\mu}+\frac{1}{2N}+\frac{1}{4N^2}
\end{equation}
This result confirms again that $\lambda_s=\beta_s=1$, consistent with the result of path integral calculation and our numerical results (Fig.~\ref{fig:SM_eqscalingcollapse}).

\end{document}